\documentclass[twocolumn,aps,pra, floatfix,superscriptaddress]{revtex4-1}
\usepackage{amssymb,amsthm, amssymb, latexsym}
\usepackage{graphicx}
\usepackage{epstopdf}
\usepackage{amsmath}
\usepackage[english]{babel}
\usepackage{float}
 \usepackage{color}

\begin{document}
\title{Excitation and dynamics in the extended bose-hubbard model}

\author{Beno\^{\i}t~Gr\'{e}maud}

\affiliation{MajuLab, CNRS-UNS-NUS-NTU International Joint Research
  Unit UMI 3654, Singapore}

\affiliation{Centre for Quantum Technologies, National University of
  Singapore, 3 Science Drive 2, Singapore 117543, Singapore}

\affiliation{Department of Physics, National University of Singapore,
  2 Science Drive 3, Singapore 117542, Singapore}

\affiliation{Laboratoire Kastler Brossel, UPMC-Sorbonne Universit\'es,
  CNRS, ENS-PSL Research University, Coll\`{e}ge de France, 4 Place
  Jussieu, 75005 Paris, France}

\author{G.G. Batrouni}

\affiliation{INLN, Universit\'e de Nice--Sophia Antipolis, CNRS; 1361
  route des Lucioles, 06560 Valbonne, France}

\affiliation{Institut Universitaire de France, 103, Boulevard
  Saint-Michel, 75005 Paris, France}

\affiliation{MajuLab, CNRS-UNS-NUS-NTU International Joint Research
  Unit UMI 3654, Singapore}

\affiliation{Centre for Quantum Technologies, National University of
  Singapore, 3 Science Drive 2, Singapore 117543, Singapore}

\date{\today}

\begin{abstract}

The one-dimensional extended bosonic Hubbard model has been shown to
exhibit a variety of phases ranging from Mott insulator and superfluid
to exotic supersolids and Haldane insulators depending on the filling
and the relative value of the contact ($U$) and near neighbor ($V$)
interaction strengths. In this paper we use the density matrix
renormalization group and the time evolving block decimation numerical
methods to study in detail the dynamics and excitation spectra of this
model in its various phases. In particular, we study in detail the
behavior of the charge and neutral gaps which characterize the Mott,
charge density and Haldane insulating phases. We also show that in
addition to the gapless modes at $k=0$, the supersolid phase exhibits
gapless modes at a finite $k$ which depends on the filling.

\end{abstract}

\maketitle

\section{Introduction}

The bosonic Hubbard model (BHM) has continued to attract interest
since its introduction by Fisher {\it et al.} \cite{fisher89}. This
interest stems from its use to understand many physical phenomena such
as the effect of disorder on superfluids and the appearance of the
compressible Bose glass phase~\cite{fisher89}, quantum phase
transitions between strongly correlated exotic phases {\it etc}.
Interest in the BHM intensified with the experimental realization of
Bose-Einstein condensates and the ability to load them in optical
lattices~\cite{greiner02}. Under experimentally realizable conditions,
these systems are described by the BHM and its
extensions~\cite{jaksch99} with highly tunable parameters and in one,
two and three dimensions which makes them ideal for studying quantum
phase transitions and exotic phases in strongly correlated systems.

Increasingly, over the last several years, the physics of strongly
correlated quantum systems has focused on the existence and properties
of unconventional phases and phase transitions.  In addition to well
studied Mott insulating behavior caused by strong on-site repulsion at
commensurate filling, extensive quantum Monte Carlo (QMC) simulations
have shown that a strong enough near neighbor repulsion can lead to
insulating incompressible density wave order (CDW) at integer and half
odd integer fillings. Topological phases, such as the Haldane
insulator which is a gapped phase characterized by a non-local
(string) order parameter~\cite{haldane83,dennijs89} can be also found
in the extended BHM~\cite{altman06,altman08,batrouni13} in one
dimension.  Finally, doping these phases can lead to phase separation
or to supersolid (SS)
phases~\cite{batrouni14,batrouni95,batrouni00,goral02,wessel05,boninsegni05,sengupta05,otterlo05,batrouni06,yi07,suzuki07,dang08,pollet10,capogrosso10}.

Even though the phase diagram of the extended BHM is now well
understood, the excitation spectra of the various ground states have
been less studied~\cite{jolicoeur93,kuhner00}, essentially because the
numerical methods providing the ground state properties, such as exact
diagonalization or QMC, become limited in the dynamical domain. More
recently, for quasi-1D systems, the extension of the density matrix
renormalization group method (DMRG) to the time domain or,
equivalently, the time evolving density matrix method (TEBD) have
proved to be extremely successful in probing the dynamical properties
of the system, thereby providing reliable excitation
spectrum~\cite{alps,tebd-vidal,Schollwock_review}.  In this paper, we
extend our work in~\cite{batrouni13,batrouni14} to study the
excitation spectrum of the one dimensional extended BHM in different
phases, namely the Mott insulating phase (MI), the Haldane insulating
phase (HI), the charge density wave phase (CDW) and the supersolid
phase (SS).

The paper is organized as follows. In section~\ref{sec:model} we
present the model and the various methods to compute the ground state
properties and excitation spectrum. In section~\ref{sec:mihicdw}, we
present the dependence of the dynamical structure factor on the near
neighbor repulsion, $V$, at fixed filling, $n=1$, and fixed value of
the contact repulsion, $U$ and hopping, $t$.  In
section~\ref{sec:supersolid}, we study the dynamical structure factor
in the SS phase for different fillings and explain its main properties
using a mapping of the extended BHM to the Heisenberg model for a spin
1/2 chain in a finite magnetic field. In addition, we describe the
evolution of the spectrum across the SS-SF transition.  In
section~\ref{sec:undercdw}, we discuss the qualitative differences
found in the dynamical structure factor in the phase which is obtained
by underdoping the half-filled CDW and in the SS phase.  A summary of
results and conclusions is in section~\ref{sec:conclusion}.

\section{Model and methods}
\label{sec:model}

\subsection{The model Hamiltonian}

The one dimensional extended BHM we shall study is described by the
Hamiltonian,
\begin{eqnarray}
\nonumber
 H &=& -t\sum_{i} (a^{\dagger}_ia^{\phantom\dagger}_{i+1} +
a^{\dagger}_{i+1}a^{\phantom\dagger}_{i}) + \frac{U}{2} \sum_i
n_i\left(n_i-1\right)\\
&& + V\sum_in_in_{i+1}.
\label{ham}
\end{eqnarray}
The sum over $i$ extends over the $L$ sites of the lattice, periodic
boundary conditions were used in the QMC simulations~\cite{sgf} and
open conditions in the DMRG and the TEBD simulations.  The onsite
repulsive interaction energy, $U$, is put equal to unity and sets the
energy scale. The time scale being given by $\hbar/U$, it is also put
equal to unity. The operator $a_i^{\phantom\dagger}$ ($a_i^\dagger$)
destroys (creates) a boson on site $i$, $n_i= a_i^\dagger
a_i^{\phantom\dagger}$ is the number operator on site $i$, $t$ is the
hopping amplitude and $V$ is the near neighbor repulsive interaction
parameter. Since we will typically study the system in the canonical
ensemble, we did not include a chemical potential term in $H$.

The charge gap is given by,
\begin{eqnarray}
\label{chargegapmu}
\Delta_c(n)&=&\mu(n)-\mu(n-1)\\
&=& E_0(n+1)+E_0(n-1)-2E_0(n)
\label{chargegapE}
\end{eqnarray}
where the chemical potential is given by $\mu(n)=E_0(n+1)-E_0(n)$ and
$E_0(n)$ is the ground state energy of the system with $n$ particles
and is obtained both with QMC and DMRG. The neutral gap, $\Delta_n$,
is obtained using DMRG by targeting the lowest excitation with the
same number of bosons. For the CDW and HI phases, the chemical
potentials at both ends are set to (opposite) large enough values,
when using DMRG, such that the ground state degeneracy and the low
energy edge excitations are
lifted~\cite{kuhner00,altman06,rossini12,batrouni13}.

For a bosonic filling $\bar{n}$ close to unity, the Bose-Hubbard model
can be reasonably approximated by the AF spin-1 Heisenberg model:
\begin{equation}
\label{Heisenberg}
H_S=\sum_{i}J(S_{i}^xS_{i+1}^x+S_{i}^yS_{i+1}^y)+\lambda
S_{i}^zS_{i+1}^z+D\left(S_{i}^z\right)^2,
 \end{equation}
 where $\lambda$ is the axial anisotropy and $D$ the ion
 anisotropy. One has the following mapping between the
 parameters~\cite{altman08}: $J=-t\bar{n}$, $\lambda=V$ and $D=U/2$.

\subsection{Time Evolving Block Decimation}

As mentioned above, the excitation spectra are obtained using the
TEBD, first in imaginary time to obtain the ground state, then in real
time to compute the density-density correlation function. In each
case, we have used a number preserving algorithm. We have checked that
the ground state properties (energy, site density, double
occupancy...)  obtained with the TEBD exactly match the properties of
the ground state obtained from the DMRG, using the ALPS
library~\cite{alps}. The space and time correlation functions $\langle
A_i(T)B_j\rangle$, where $\langle\cdots\rangle$ is the ground state
average and where $A_i(T)$ is the time evolution of the operator $A_i$
in the Heisenberg picture, have been obtained by writing,
\begin{equation}
\label{LR}
\langle A_i(T)B_j\rangle=e^{iE_{GS}T}\langle GS|A_i e^{-iHT}B_j|GS\rangle,
\end{equation}
where $E_{GS}$ is the ground state energy. Therefore, computing the
correlation function can be done as follows:
\begin{itemize}
\item From the matrix product state (MPS) representation of the ground
  state, $|GS\rangle$, one computes the MPS of the initial state
  $|\Psi(0)\rangle=B_j|GS\rangle$.

\item The state $|\Psi(0)\rangle$ is evolved using the real time TEBD,
  providing the MPS of $|\Psi(T)\rangle$, and thereby allowing the
  computation of the correlation function as $\langle
  GS|A_i|\Psi(T)\rangle$.
\end{itemize}

In what follows, we focus on the density-density correlation,
i.e. $A_i=B_i=n_i-\langle n_i\rangle$, more precisely, the initial
state consists of creating a density excitation in the middle of the
chain: $B_j|GS\rangle$ with $j=L/2$. We then compute $\langle
A_i(T)B_j\rangle$ for all sites and times $T$ up to $150$, with a time
step equal to $0.1$ (a smaller timestep was actually used for the
propagation).  Finally, the
dynamical structure factor $S(k,\omega)$ is computed from the Fourier
transform of the density-density correlation with respect to $i-j$ and
$T$. In order to smooth out oscillations caused by the finite time
window, we actually compute the Fourier transform of $\langle
A_i(T)B_j\rangle \exp{(-4T^2/T^2_{\text{max}})}$.

\section{Mott-Haldane-CDW transitions}
\label{sec:mihicdw}

The qualitative description of the different phases is based on the
Heisenberg model EQ.\eqref{Heisenberg}. More precisely, the phases are
characterized by the values of the string order parameters:
\begin{equation}
  O^{\alpha}=\lim_{|i-j|\rightarrow\infty} \langle
  S^{\alpha}_ie^{i\pi\sum_{p=i+1}^{j-1}S^{\alpha}_p}S^{\alpha}_j\rangle.
\end{equation}
characterizing a loose antiferromagnetic order along the different
axes $\alpha=x,y,z$. They are associated with an underlying non-local
discrete $Z_2\otimes Z_2$ symmetry of the Heisenberg model,
Eq.\eqref{Heisenberg}~\cite{Affleck87}. In the large-$D$ phase (the MI
state for bosons), the $O^{\alpha}$ vanish. In the Haldane phase, the
two discrete $Z_2$ symmetries are broken, resulting in nonvanishing
string order $O^{\alpha}$.  Finally, in the Ising phase (the CDW for
bosons), only the $Z_2$ symmetry along the $z$-axis is broken such
that only the string order $O^{z}$ is non-vanishing. Note that in that
phase, the string order and the antiferromagnetic order (the CDW order
for bosons) are equivalent~\cite{Oshikawa92}.

The lowest elementary excitations from the ground state ($ka=0$,
$S_z=0$) are part of a triplet, one neutral $\epsilon^{(0)}(k)$
($\delta N=S_z=0$,), two charge ones $\epsilon^{(\pm)}(k)$ ($\delta
N=S_z=\pm1$)~\cite{jolicoeur93, fath93}, where $\delta N$ corresponds
to change in the total number of bosons. In each sector, one defines a
gap which corresponds to the minimum of the elementary excitations
over all $k$ values: $G^{(0)}$ and $G^{(\pm)}$.

The minima are located either at $ka=0$ or $ka=\pi$.  From linear
response theory, the structure factor reads
\begin{equation}
 S(k,\omega)\propto \sum_m\frac{|\langle
   k,m|\delta\hat{n}|GS\rangle|^2}{\omega+i\eta+E_{GS}-E_{k,m}},
\end{equation}
where $|k,m\rangle$ denotes the different excited states of $H$ for a
given momentum $k$, and $E_{k,m}$ the corresponding energy. For single
excitations, one simply has $E_{k,m}=\epsilon^{(m)}(k)$. Doubly
excited states for fixed $k$ are made of two single excitations:
$|q,m;k-q,m'\rangle$, corresponding to an energy
$\epsilon^{(m)}(q)+\epsilon^{(m')}(k-q)$.

By definition, the charge gap of the system is $\Delta_c= G^{(+)}+
G^{(-)}$, i.e. the minimum energy for adding a particle plus the
minimum energy for removing a particle (increasing or decreasing
$S_z$, in the Heisenberg model).  The neutral gap corresponds to the
minimum of either the elementary neutral excitations, i.e. $G^{(0)}$,
or of $\epsilon^{(+)}(k-q)+\epsilon^{(-)}(q)$, i.e., a combination of
two charge excitations.  Since, the minimum of both the elementary
charge excitation $\epsilon^{(\pm)}(q)$ is attained at either $q=0$ or
$q=\pi$, the minimum of the two-particle excitation necessarily takes
place at $k=0$ and has the value $G^{(+)}+G^{(-)}$, and corresponds
then to the lower bound of a two-particle continuum.  In short, the
neutral gap value is given by the minimum of $G^{(0)}$ and
$G^{(+)}+G^{(-)}$.

\begin{figure}[H]
\centerline{\includegraphics[width=8cm]{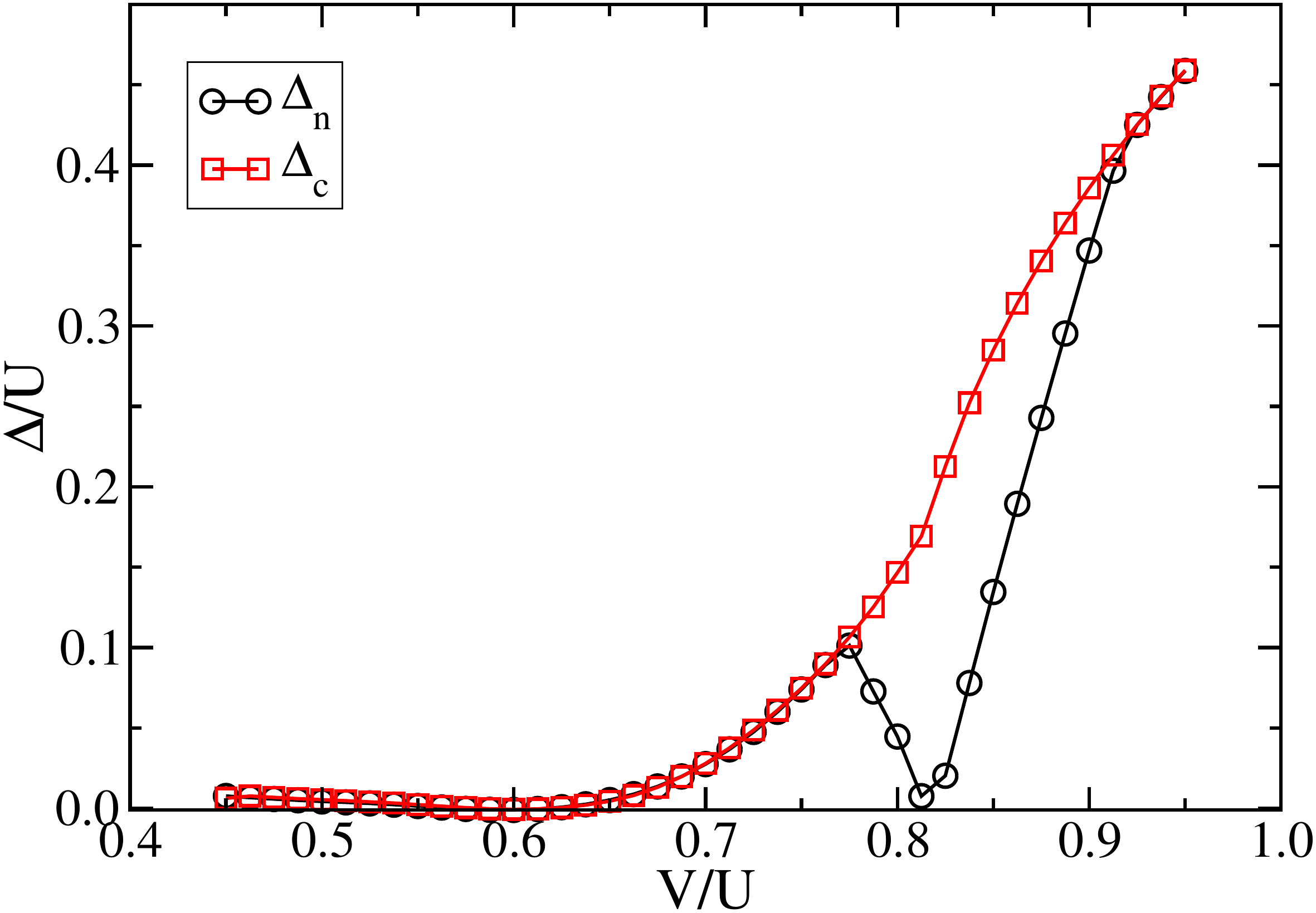}}
\caption{\label{MHC}Mott - Haldane - CDW transitions at fixed
  $t/U=0.25$. Around $V_C\approx0.75U$, the neutral and charge gaps
  start to differ, indicating that $G^(0)<G^{(+)}+G^{(-)}$: the gap
  for the single particle neutral excitations is smaller than the gap
  for the two particle excitations. The fact that only the neutral gap
  vanishes at the HI-CDW transition is a signature that the string
  order $O^{z}$ along the $z$-axis remains finite, where $O^{x}$ and
  $O^{y}$ orders vanish. }
\end{figure}

It is well-known that in the Haldane phase, the neutral gap changes
from one type to the other~\cite{jolicoeur93, fath93,altman06} and is
emphasized in Fig.~\ref{MHC}, where around $V_C\approx0.75U$, the
neutral and the charge gaps start having different values. For lower
$V$ values, one has $\Delta_n=\Delta_c=G^{(+)}+G^{(-)}$, whereas for
larger $V$ values, one has
$G^{(0)}=\Delta_n<\Delta_c=G^{(+)}+G^{(-)}$.  This results from the
fact that, in the Haldane phase, the elementary excitations are all
gapped, with a minimum occuring at $ka=\pi$~\cite{jolicoeur93,
  fath93}. For a fixed value of $U$, corresponding to a fixed value of
the ion anisotropy, $D$, in the corresponding spin Hamiltonian,
$G^{(\pm)}$ increases with increasing $V$ (i.e. $\lambda$), whereas
$G^{(0)}$ decreases. This can be understood by starting at the
Heisenberg point, ($D=0$, $\lambda=J$), where, due to $SU(2)$
invariance of the spin Hamiltonian, all single particle excitation
energies are the same, so that $G^{(0)}=G^{(\pm)}$. Increasing the ion
anisotropy, $D$, i.e. going toward the Mott Phase (or decreasing
$\lambda$) gives rise to a smaller in-plane gap (i.e. the elementary
charge gap) $G^{(\pm)}<G^{(0)}$.

This evolution of the neutral and charge gaps can be seen in the
behavior of the structure factor $S(k,\omega)$. We emphasize that even
though the structure factor $S(k,\omega)$ is a neutral excitation,
i.e. conserves the total number of bosons, it also couples to the
two-particle continuum composed of elementary charge excitations
$\epsilon^{(+)}(k-q)+\epsilon^{(-)}(q)$. As explained above, in the
limit $k\rightarrow0$, the minimum energy corresponds to the charge
gap $\Delta_c=G^{(+)}+G^{(-)}$, such that even if we expect
$S(k,\omega)$ to vanish at $ka=0$, the value for $\Delta_c$ can be
obtained by extrapolating the behavior of $S(k,\omega)$ around $ka=0$.

\begin{figure}[H]
\includegraphics[width=4.0cm]{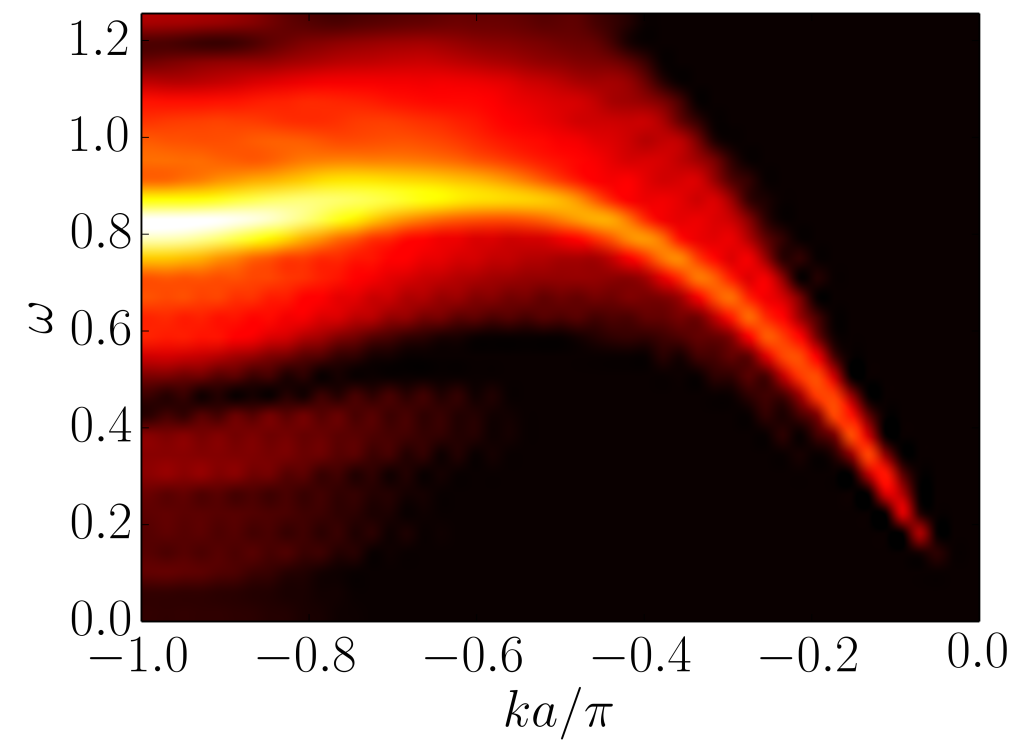}
    \includegraphics[width=4.0cm]{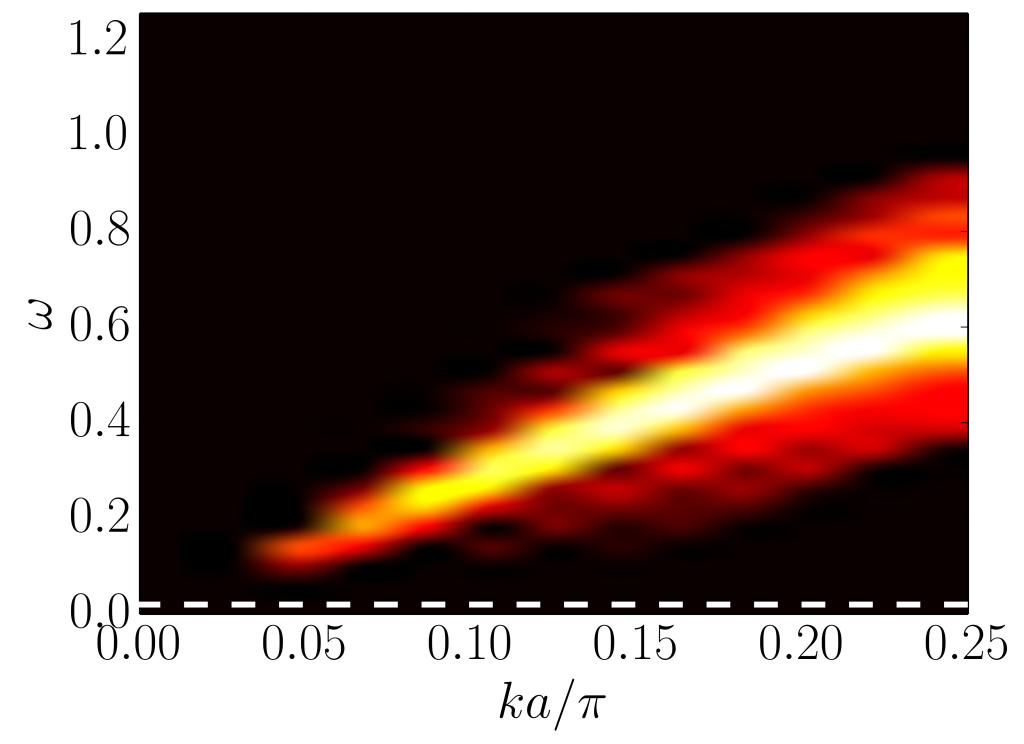}\\
\includegraphics[width=4cm]{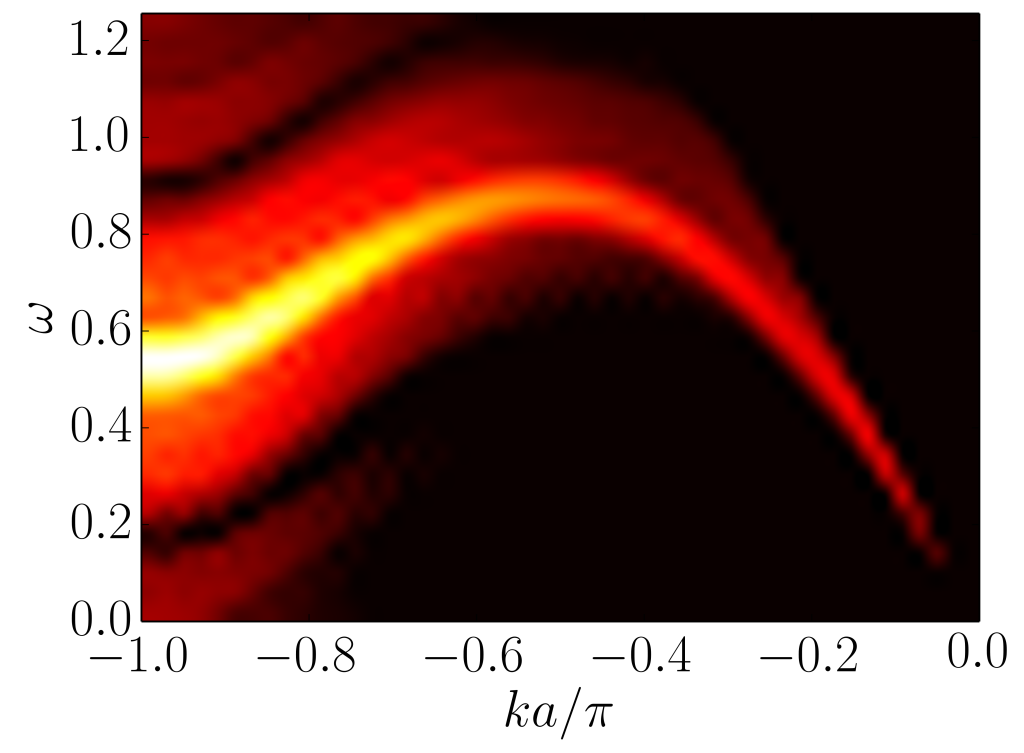}
    \includegraphics[width=4cm]{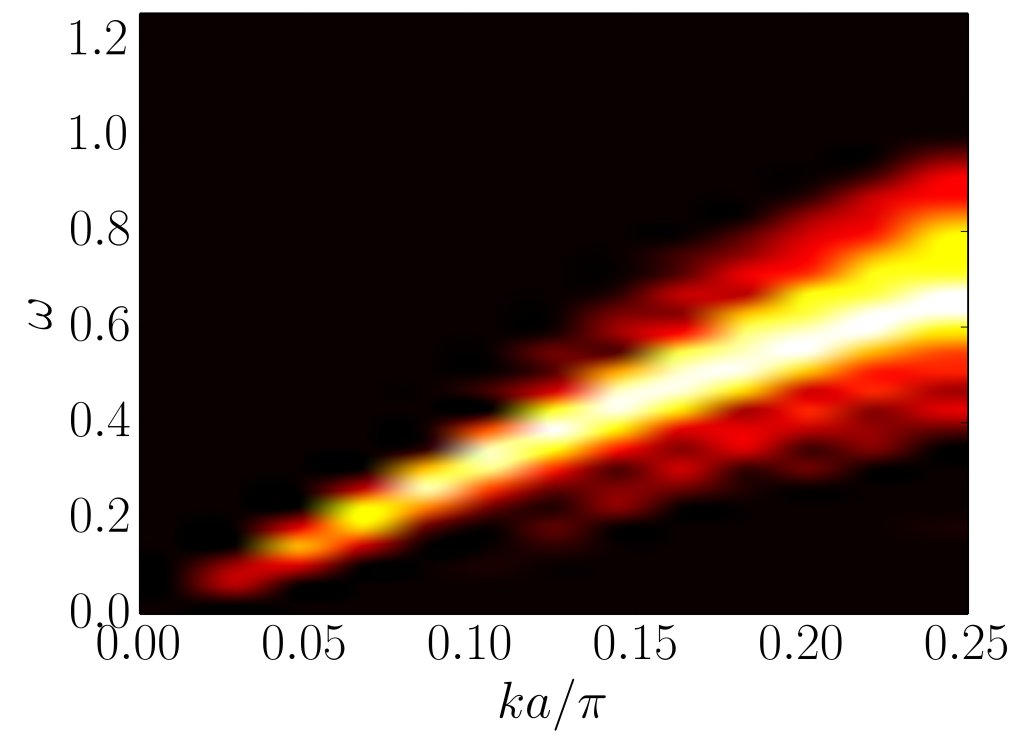}
    \caption{\label{MI} TEBD excitation spectra. TOP: $V/U=0.4$,
      $t/U=0.25$ MI phase. The gap at $ka=\pi$ is much larger than the
      gap at $ka=0$.  $\Delta_c=\Delta_n= S(k\to 0,\omega)$. The white
      dashed line in the right panel shows the value of the gap (small
      but non-vanishing). BOTTOM: $V/U=0.58$, $t/U=0.25$ at the MI-HI
      transition. The excitations are almost gapless near $ka=0$,
      whereas the excitation is cleary gapped at $ka=\pi$. See text.}
\end{figure}

In the Mott phase, top Fig.~\ref{MI}, one can clearly see that the gap
at $ka=\pi$ is much larger than the gap at $ka=0$. Note that the gap
$ka=\pi$ differs from $G^{(0)}$, since, in the MI phase,
$\epsilon^{(0)}(k)$ is minimum at $ka=0$ and maximum at $ka=\pi$.  The
neutral and the charge gap have the same value $G^{(+)}+G^{(-)}$,
which can be obtained from $S(k,\omega)$ by extrapolating the gap
value to $ka=0$. At the Mott-Haldane transition, bottom Fig.~\ref{MI},
$S(k,\omega)$ exhibits (almost) gapless excitation around $ka=0$,
whereas the excitation is cleary gapped at $ka=\pi$. Since the
transition corresponds to breaking both hidden $Z_2$ symmetries, both
the neutral and the charge gaps vanish, corresponding to vanishing
elementary charge excitations gap $G^{(\pm)}$, but a finite elementary
neutral excitation gap $G^{(0)}$.

\begin{figure}[H]
\hskip 0.25cm\includegraphics[width=4cm]{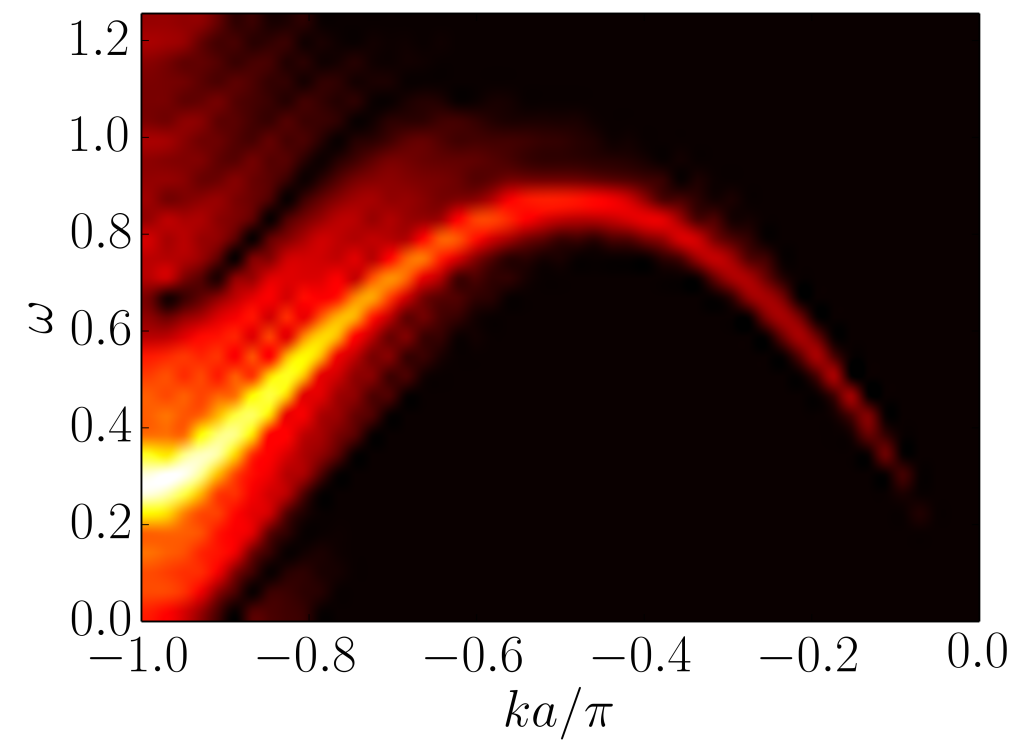}
    \includegraphics[width=4cm]{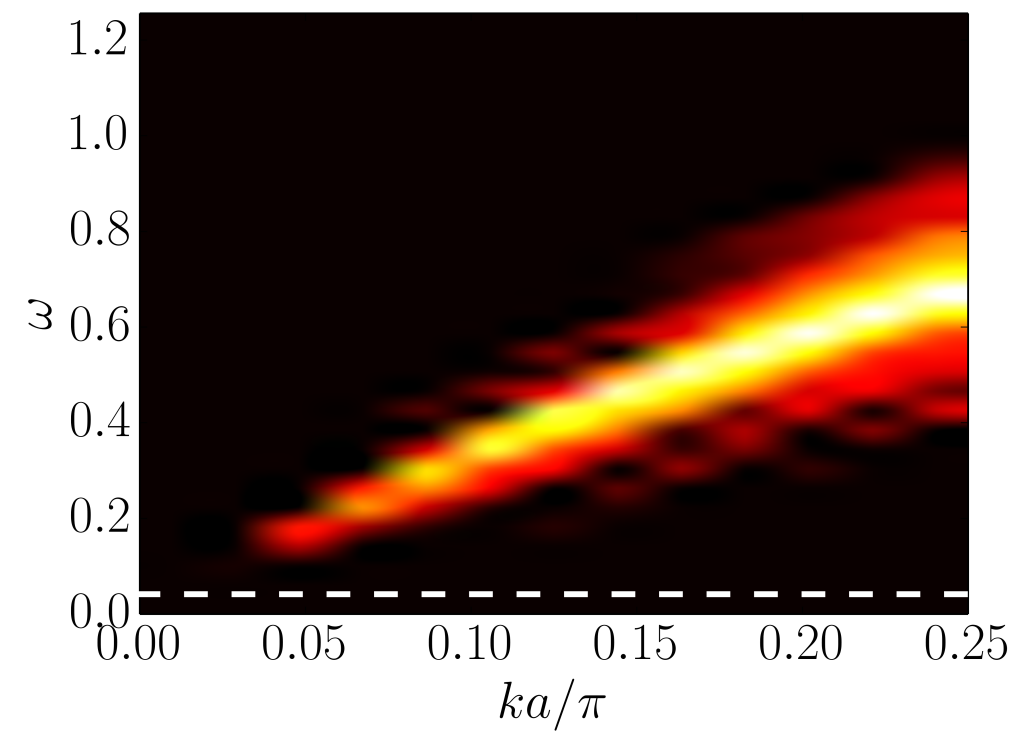}\\
  \centerline{\includegraphics[width=4cm]{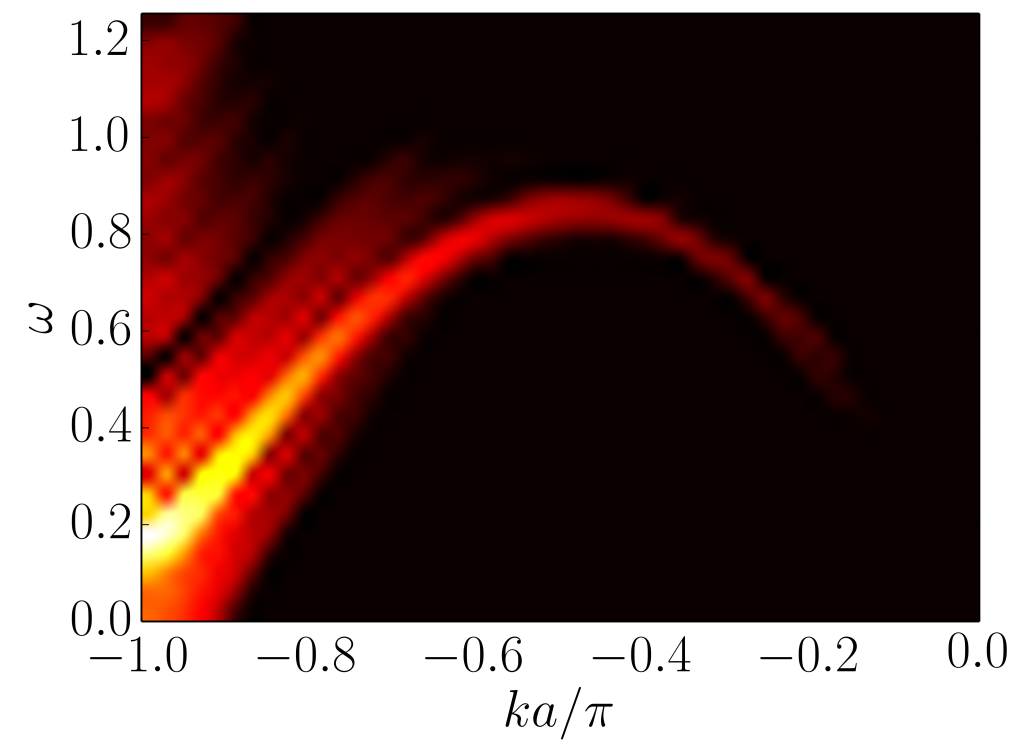}
    \includegraphics[width=4cm]{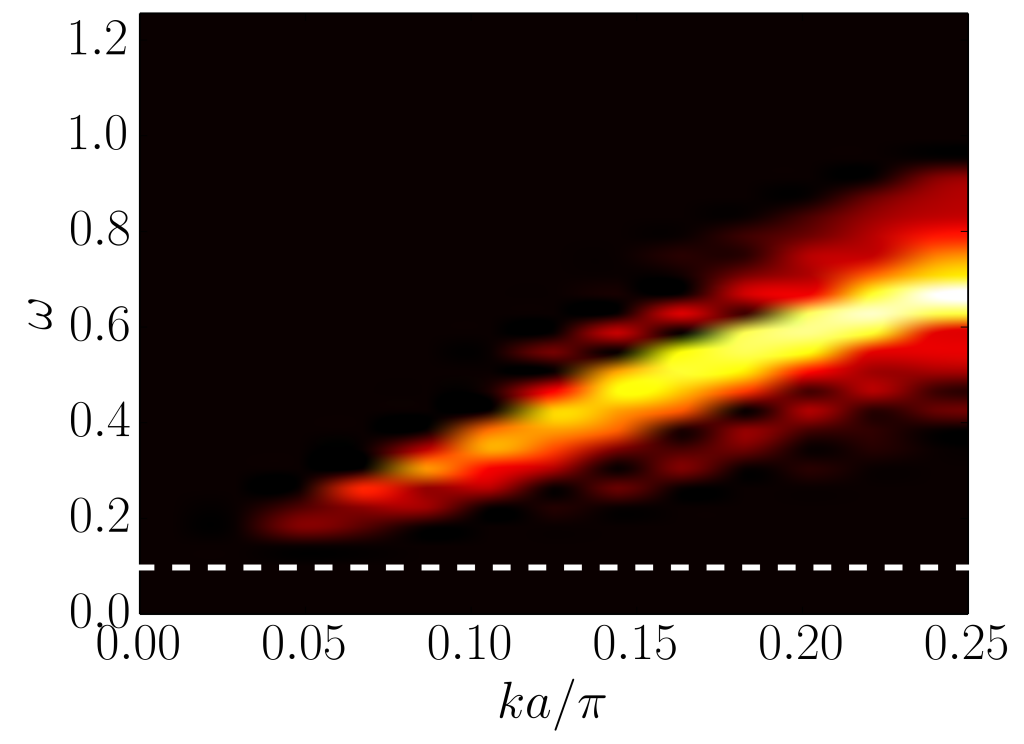}}\\
  \centerline{\includegraphics[width=4cm]{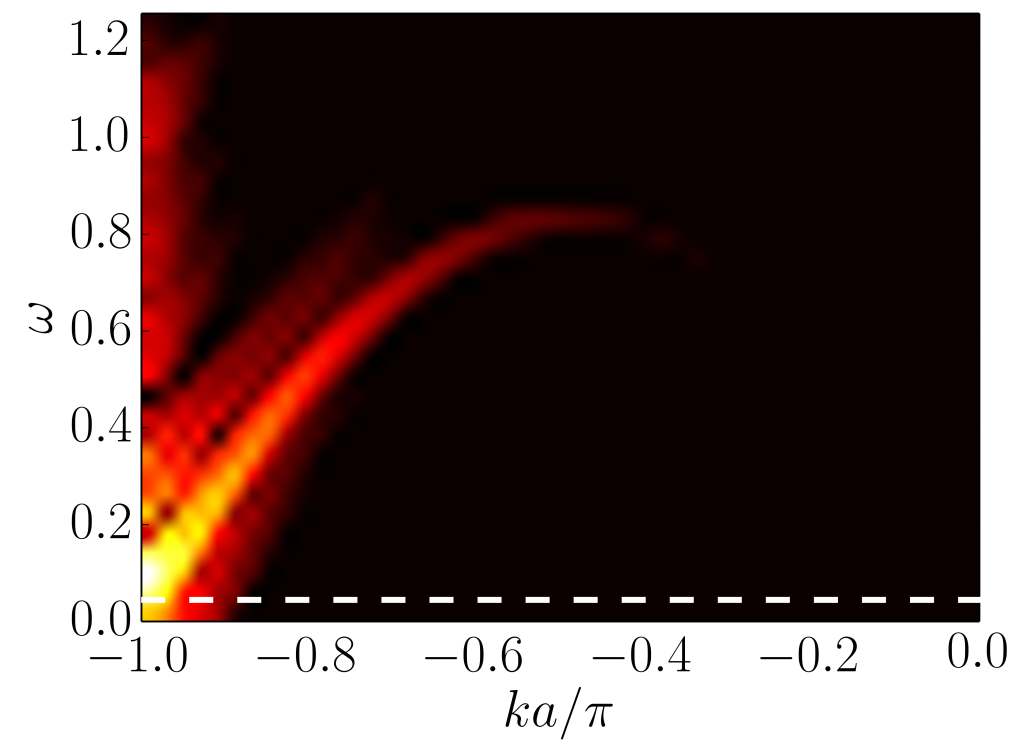}
    \includegraphics[width=4cm]{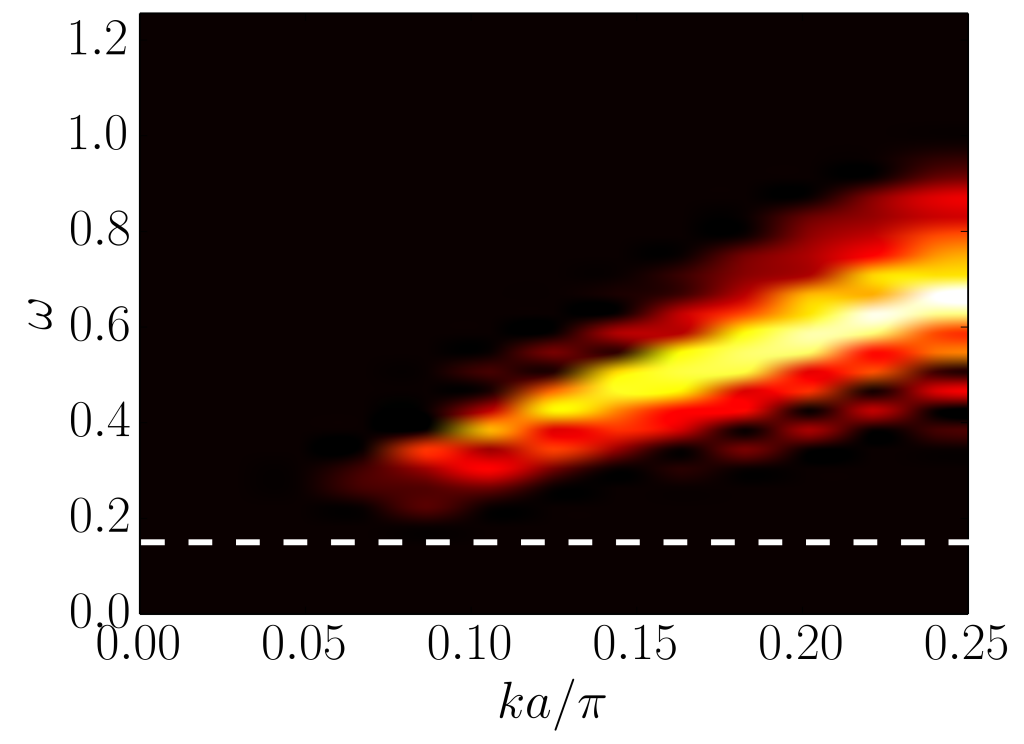}}
  \caption{\label{HIC}TEBD excitation spectra inside the HI
    phase. TOP: $V/U=0.7$, $t/U=0.25$, the gap at $ka=\pi$,
    i.e. $G^{(0)}$, decreased while the gap at $ka=0$ is nonvanishing,
    $\Delta_n=\Delta_c=G^{(+)}+G^{(-)} \neq 0$.  The white dashed line
    gives the value of the gap. MIDDLE: $V/U=0.75$, $t/U=0.25$ inside
    the HI phase at the symmetric point. The gap at $ka=\pi$,
    $G^{(0)}$, has the same value as the gap at $ka=0$,
    $G^{(+)}+G^{(-)}$, indicated by the white dashed line. BOTTOM:
    $V/U=0.79$, $t/U=0.25$ inside the HI phase. The gap at $ka=\pi$,
    $\Delta_n=G^{(0)}$, is smaller than the charge gap at $ka=0$,
    $\Delta_c=G^{(+)}+G^{(-)}$.  The white dashed line in the left
    plot corresponds $\Delta_n=G^{(0)}$, whereas, in the right plot,
    it corresponds to $\Delta_c=G^{(+)}+G^{(-)}$.  }
\end{figure}

Inside the Haldane phase, but for a value $V<V_c$ , top
Fig.~\ref{HIC}, we see that the gap at $ka=\pi$, i.e. $G^{(0)}$, has
decreased whereas the gap at $ka=0$, i.e. $\Delta_n=\Delta_c =
G^{(+)}+G^{(-)}$, is non vanishing. At the value $V\approx V_c$,
middle Fig.~\ref{HIC}, the gap at $ka=\pi$, $G^{(0)}$, has almost the
same value as the gap at $ka=0$, $G^{(+)}+G^{(-)}$. When $V>V_c$,
bottom Fig.~\ref{HIC}, we see that one is now in the opposite
situation: $\Delta_n=G^{(0)}$ is clearly smaller than
$\Delta_c=G^{(+)}+G^{(-)}$.

\begin{figure}[H]
\includegraphics[width=4cm]{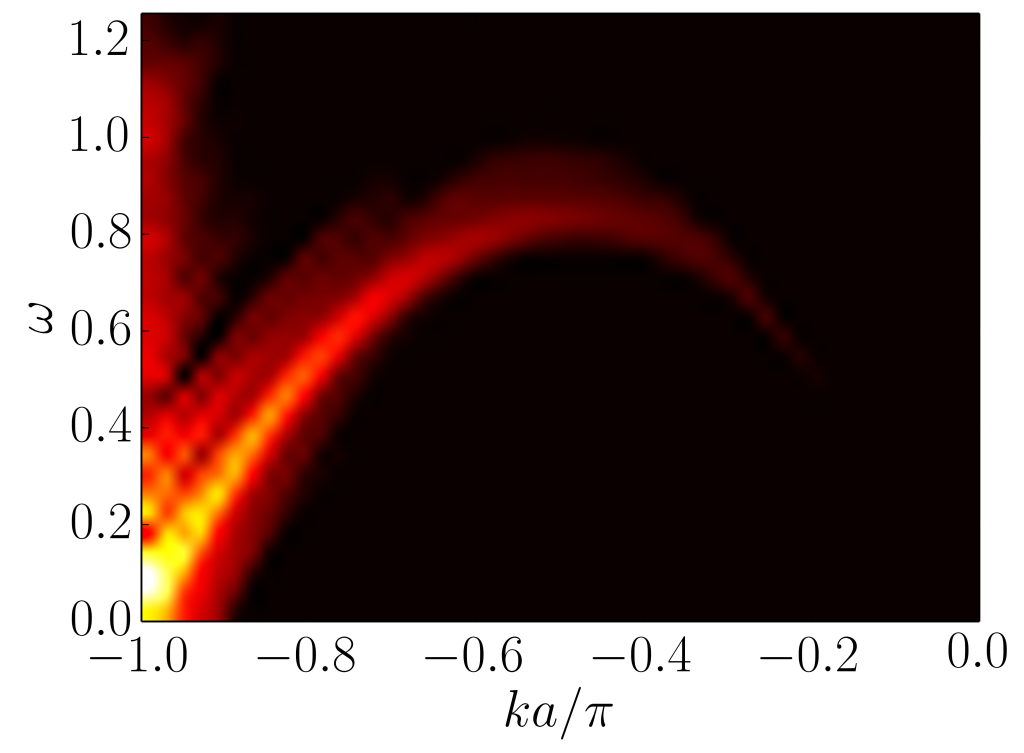}
\includegraphics[width=4cm]{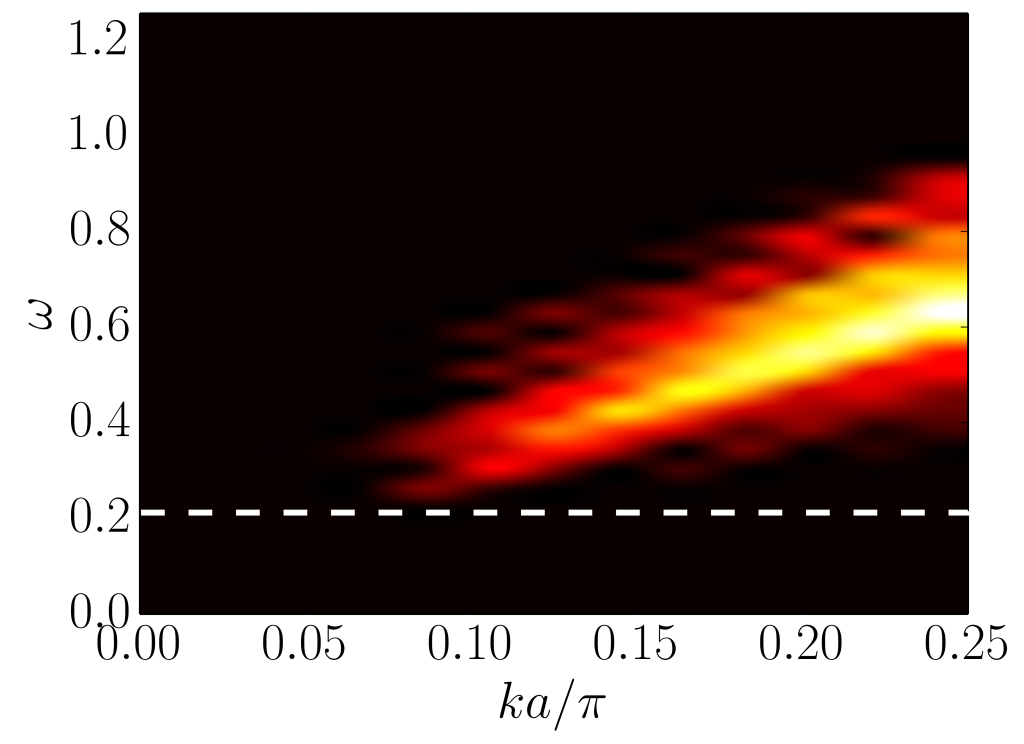}\\
\includegraphics[width=4cm]{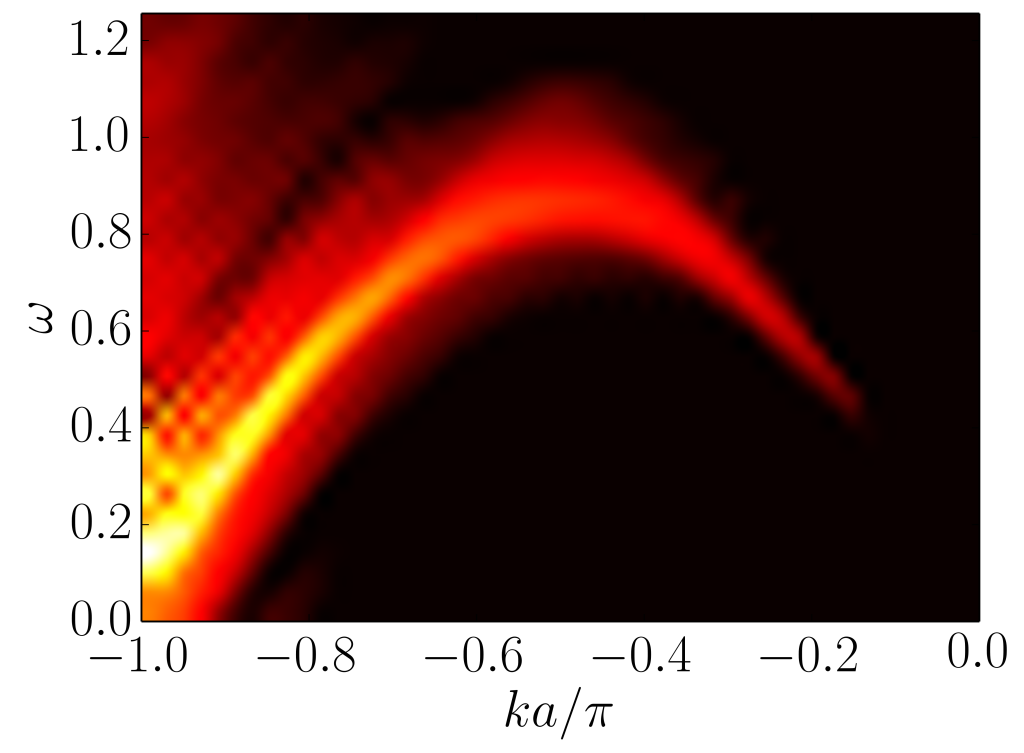}
\includegraphics[width=4cm]{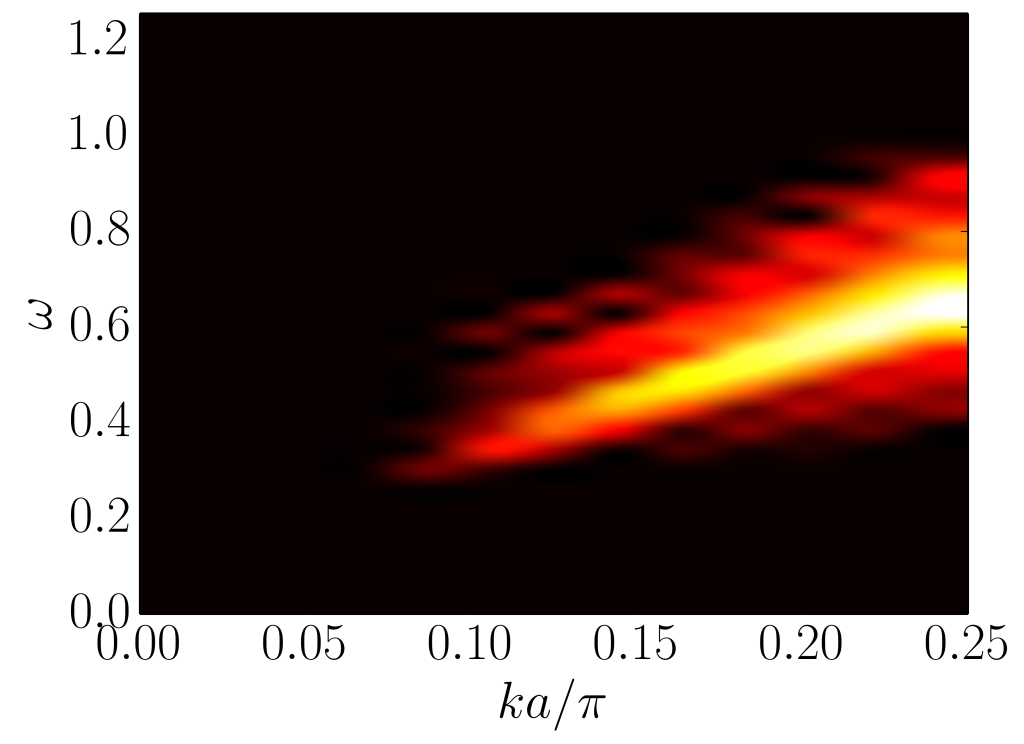}
  \caption{\label{HICDW}TEBD excitation spectra. TOP: $V/U=0.82$,
    $t/U=0.25$ at the HI-CDW transition, At $ka=\pi$,
    $\Delta_n=G^{(0)} = 0$ whereas $\Delta_c=G^{(+)}+G^{(-)}$ at
    $ka=0$, remains finite and is given by the white dashed line in
    the right plot. BOTTOM: $V/U=0.84$, $t/U=0.25$ in the CDW phase,
    the neutral gap at $ka=\pi$ is finite, but has smaller value than
    the charge gap: $\Delta_n=G^{(0)}<G^{(+)}+G^{(-)}=\Delta_c$.}
\end{figure}

At the transition between the Haldane phase and the charge density
wave phase, top Fig.~\ref{HICDW}, $\Delta_n=G^{(0)}$ vanishes while
the charge gap at $ka=0$, $G^{(+)}+G^{(-)}$ remains finite. This
corresponds to the fact that across the transition, the hidden $Z_2$
symmetry along the $z$-axis remains broken, whereas the hidden $Z_2$
symmetry along in the $XY$ plane is restored.  Since the string order
$O^{z}$ is broken by charge excitations but left invariant under
neutral excitations, i.e. commutes with the $S^z_i$ operators, only
the charge gap is protected by the finite value of the order $O^{z}$
and remains finite at the transition. On the other hand, since the
string order $O^{x}$, which is broken by neutral excitations
(i.e. applying $S^z_i$), vanishes at the transition, the neutral gap
has to close at the transition.
 
Finally, in the CDW phase, bottom Fig.~\ref{HICDW} both the neutral
and the charge gap increase, but still having
$\Delta_n=G^{(0)}<G^{(+)}+G^{(-)}=\Delta_c$.

\begin{figure}[H]
  \centerline{\includegraphics[width=6cm]{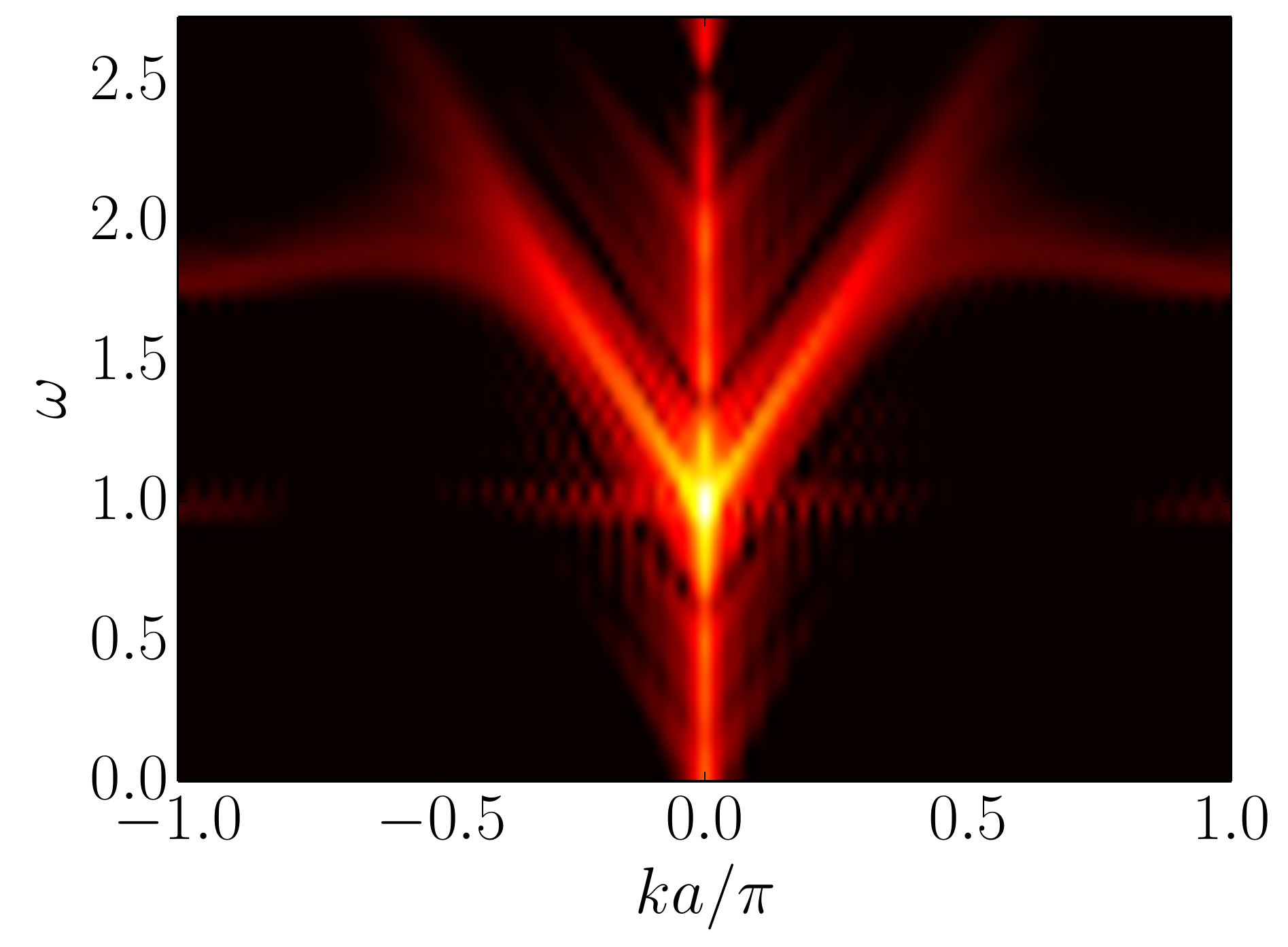}}
  \centerline{\includegraphics[width=6cm]{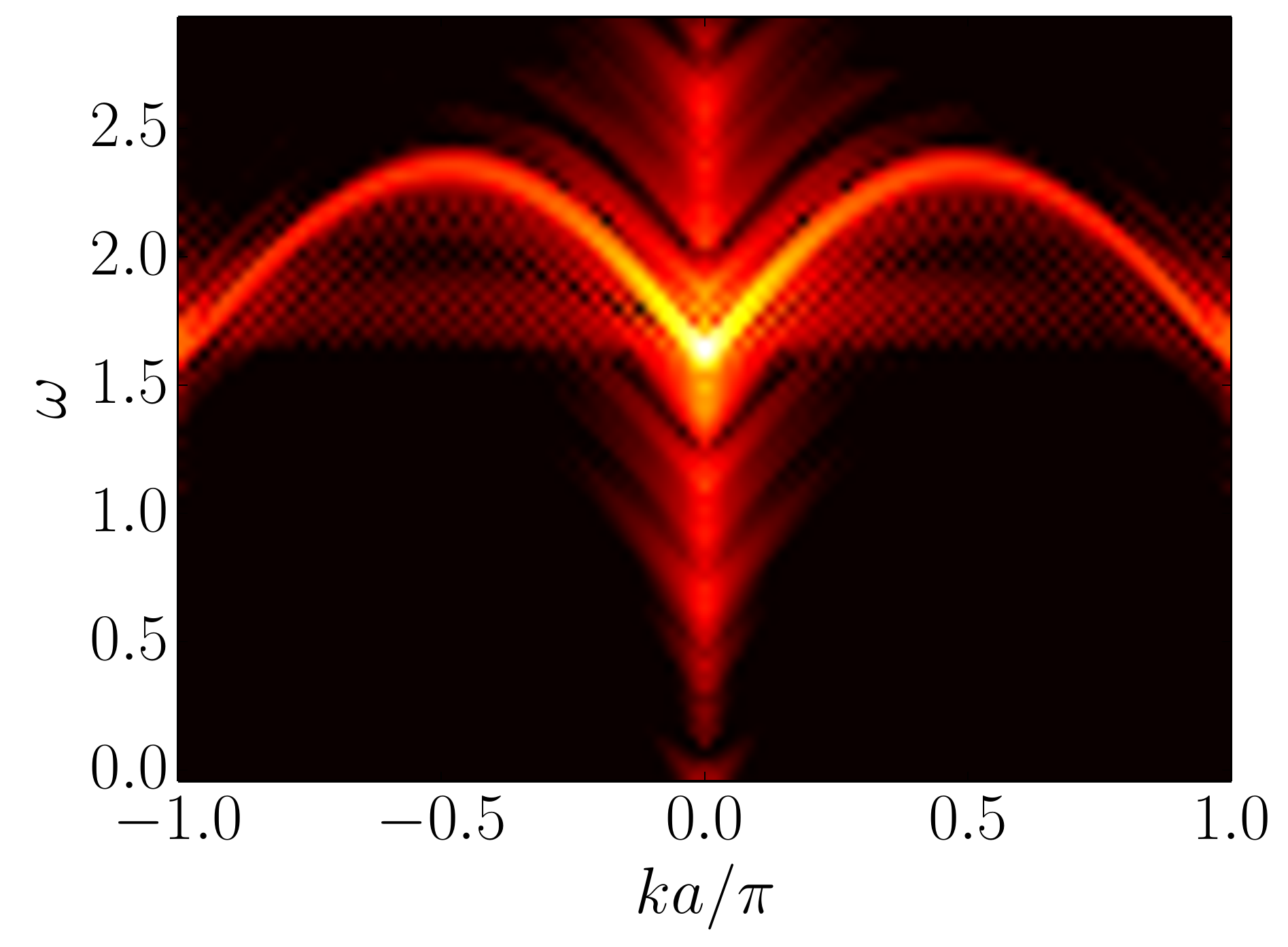}}
  \caption{\label{BD} Excitation spectra obtained when adding a boson
    to the system, i.e. corresponding to the operators $A_i=b_i$ and
    $B_j=b_j^{\dagger}$ in Eq.~\eqref{LR}. In both plots, the vertical
    offset corresponds to the chemical potential $\mu_+$ for adding a
    boson. The parameters for the top plot are the same as in
    Fig.~\ref{MI} (top panels), i.e. the Mott Phase. One clearly see
    that the minimum of the charge excitation is obtained at $ka=0$,
    the value at $ka=\pi$, being much larger. The bottom plot
    corresponds to bottom panels Fig.~\ref{HIC}, i.e. the Haldane
    phase where the neutral and the charge gap are different. The
    lowest excitation occurs at $ka=\pi$. Around $ka=0$, one has a
    two-particle continuum, made of one neutral excitation and one
    charge excitation.}
\end{figure}

Finally, in Fig.~\ref{BD}, we show the excitation spectrum when adding
a boson to the system, i.e. corresponding to the operators $A_i=b_i$
and $B_j=b_j^{\dagger}$ in Eq.~\eqref{LR}. In both plots, the large
vertical offset corresponds to the chemical potential for adding a
boson, $\mu^+$. By definition, the value of the charge gap is
$\mu^+-\mu^-$, where $\mu^-$ is the chemical potential for removing a
boson. Therefore the minimum of the excitation spectrum can be written
$\bar{\mu}+\Delta_c/2$, where $\bar{\mu}=(\mu^++\mu^-)/2$ is the
average chemical potential. The parameters for the top plot are the
same as in Fig.~\ref{MI} (top panels), i.e. the Mott Phase. One
clearly sees that the minimum of the charge excitation is obtained at
$ka=0$, the value at $ka=\pi$ being much larger. The bottom plot
corresponds to bottom panels Fig.~\ref{HIC}, i.e. the Haldane phase
where the neutral and charge gaps are different. The minimum of the
excitation occurs at $ka=\pi$, whereas around $ka=0$, one has a
two-particle continuum, made of one neutral excitation and one charge
excitation, the minimum value, at $ka=0$, being $\mu^++G^{(0)}$,
i.e. $\Delta_c/2+G^{(0)}$.

\section{Supersolid phase}
\label{sec:supersolid}

\begin{figure}[H]
  \centerline{\includegraphics[width=8cm]{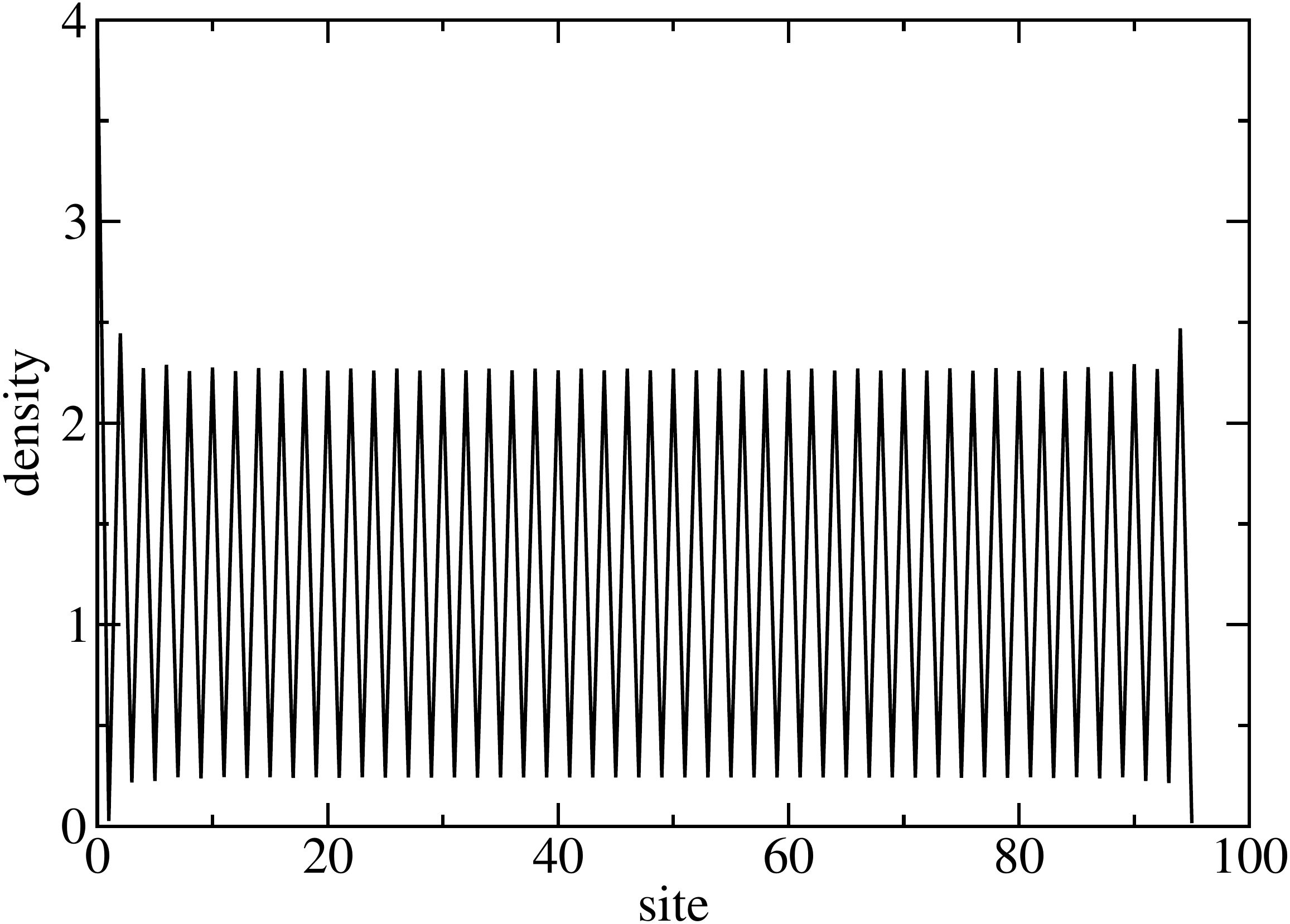}}
  \caption{\label{SSPRO}Density profile in the supersolid phase $U=1$,
    $V=0.75$ and $t=0.2$, $n=1.25$. The density oscillates between,
    $0.25$ and $2.25$, indicating long range density order at the
    incommensurate average density $n=1.25$.}
\end{figure}

The hallmark of the supersolid phase is the presence of both a long
range diagonal (density) order and superfluidity. A typical density
profile, obtained using DMRG for $U=1$, $V=0.75$ and $t=0.2$, is shown
in Fig.~\ref{SSPRO}. The oscillations of the density between $0.25$
and $2.25$, around the average value $n=1.25$ signal long range
density order but which, nonetheless, is not in the CDW phase since
the average density, $n=1.25$, is not commensurate.

\begin{figure}[H]
  \centerline{\includegraphics[width=6cm]{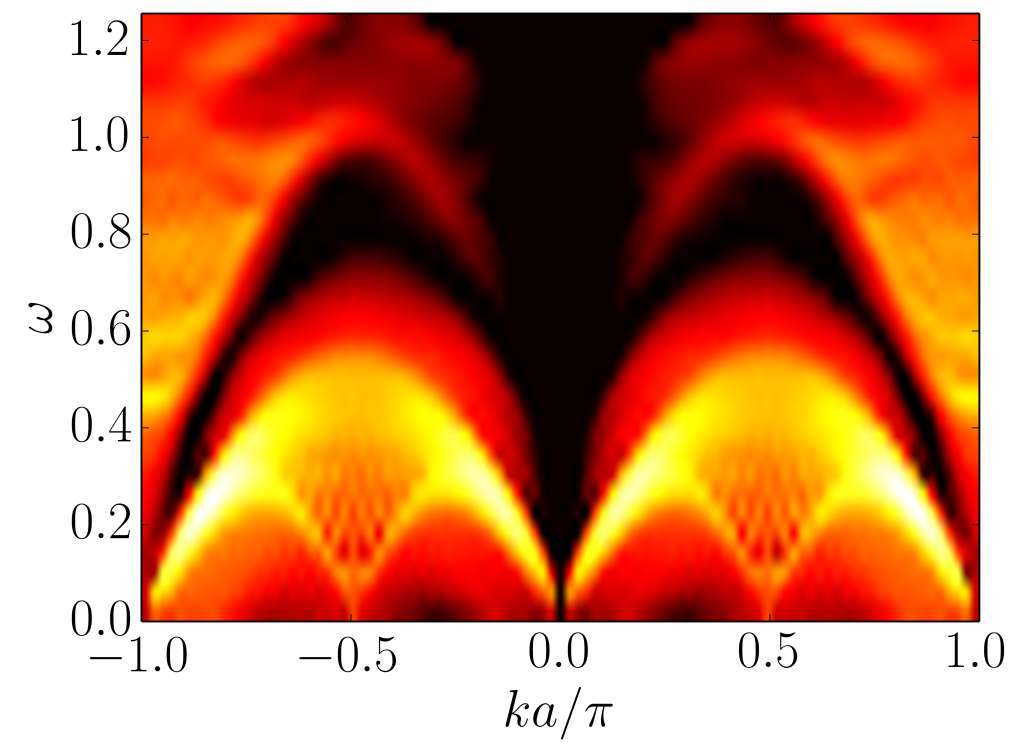}}
  \caption{\label{SS1}$t=0.2$, $n=1.25$.  The gapless excitations at
    $ka=0$ emphasize the superfluid nature of the phase.  The lower
    part of the excitation spectrum has a periodicity $\pi/a$
    reflecting the $2a$ periodicity of the low energy effective
    hamiltonian, which is an AF spin 1/2 chain (see text). The
    additional gapless mode at $ka=0.5\pi$ corresponds to the gapless
    mode at $2ka=\pi$ of the effective AF chain.}
\end{figure}

\begin{figure}[H]
  \centerline{\includegraphics[width=6cm]{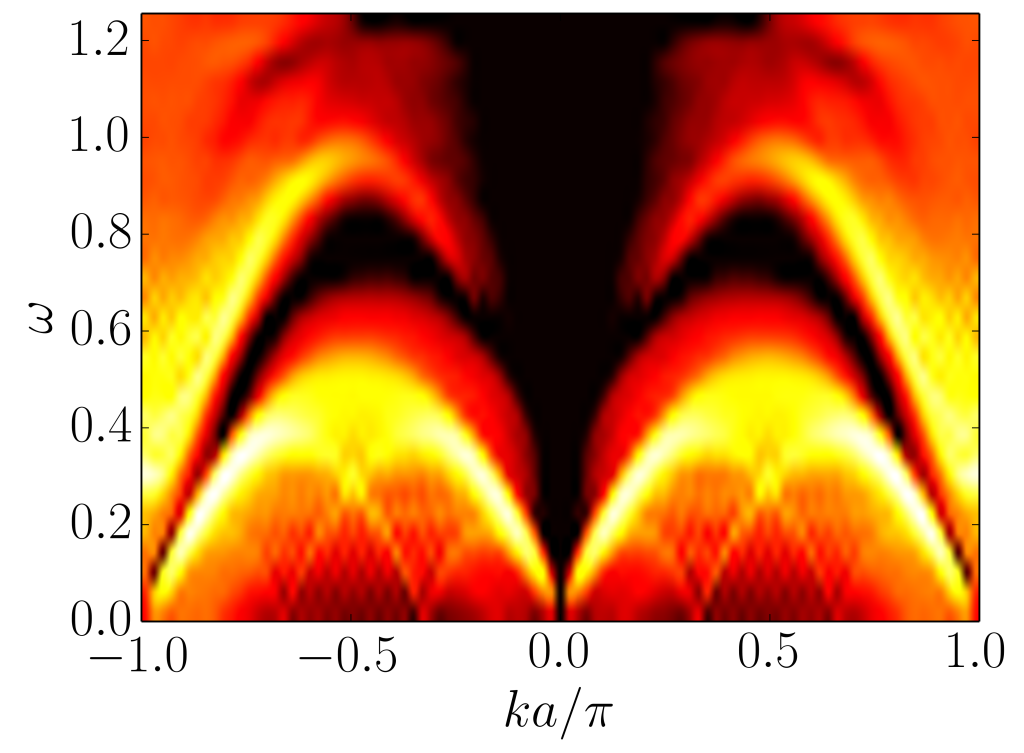}}
  \caption{\label{SS2}$t=0.2$, $n=1.167$. As in Fig.~\ref{SS1}, the
    periodicity of the lower part of the excitation spectrum can be
    understood from the low energy effective hamiltonian, which is an
    AF spin 1/2 chain in a finite magnetic field.  The additional
    gapless mode at $ka=\pi/3$ corresponds to the low energy
    incommensurate modes of the spin chain at a finite magnetization.}
\end{figure}

\begin{figure}[H]
  \centerline{\includegraphics[width=6cm]{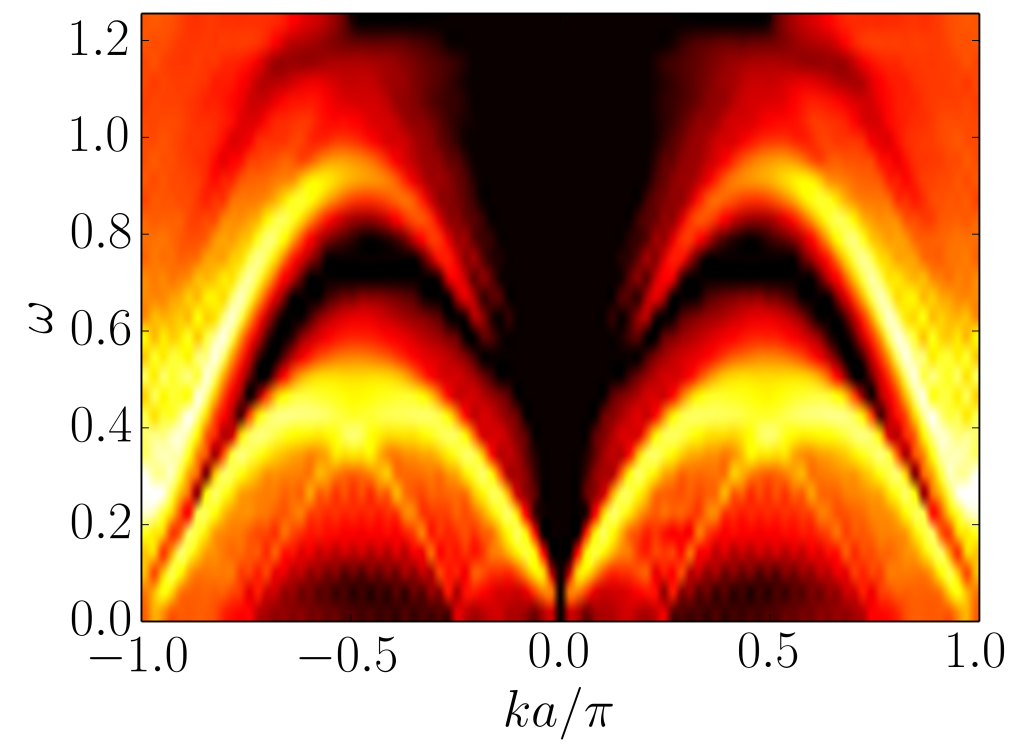}}
  \caption{\label{SS3}$t=0.2$, $n=1.125$. As in Fig.~\ref{SS2}, The
    additional gapless mode at $ka=\pi/4$ corresponds to the low
    energy incommensurate modes of a spin-1/2 chain at a finite
    magnetization.}
\end{figure}

\begin{figure}[H]
  \centerline{\includegraphics[width=6cm]{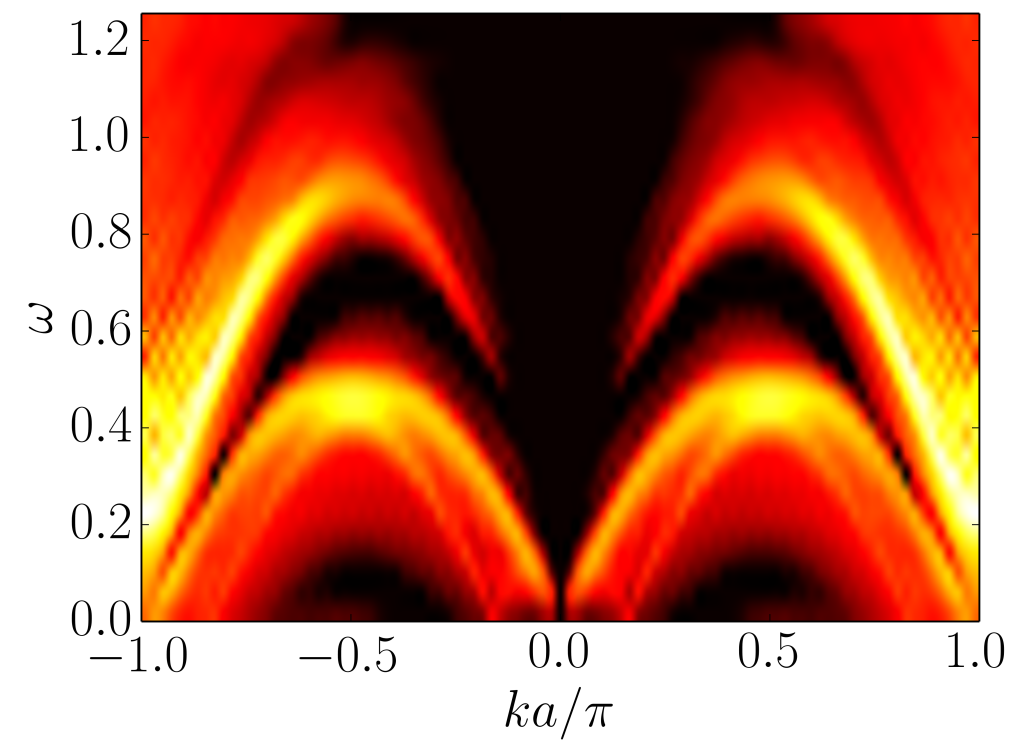}}
  \caption{\label{SS4}$t=0.2$, $n=1.08333$. As in Fig.~\ref{SS2}, The
    additional gapless mode at $ka=\pi/6$ corresponds to the low
    energy incommensurate modes of a spin-1/2 chain at a finite
    magnetization. }
\end{figure}

Figures~\ref{SS1}, \ref{SS2}, \ref{SS3} and \ref{SS4} show the the
dispersion for several values of the doping: $n=1.25$, $n=1.167$,
$n=1.125$ and $n=1.08333$ respectively. All data were obtained for
$L=96$ sites, $U=1$, $V=0.75$, $t=0.2$ and the system is in the
supersolid phase.

As expected in the supersolid phase, the system exhibits gapless
excitations at $k=0$ and $k=\pi$, but one clearly sees additional
gapless excitations at a momentum $k_{SS}$ that depends on the
density. It turns out that the value of $k_{SS}$ is in excellent
agreement with the value $2\pi\delta n$ (see below), where
$\delta_n=n-1$, i.e. $ka=\pi/2$ ($ka=\pi/3$, $ka=\pi/4$, $ka=\pi/6$ )
for $\delta_n=1/4$ ($\delta_n=1/6$, $\delta_n=1/8$, $\delta_n=1/12$).

\subsection{Effective Spin-$1/2$ Heisenberg model of the supersolid}

We consider the situation where the supersolid phase has a density
$n=5/4$, see Fig.~\ref{SSPRO}.  This occurs when one dopes the system
in the CDW phase, i.e. for $V>U/2$. The density pattern, in the limit
$U\gg t$, obtained from both the DMRG and the QMC computations show
that the ground state has a (nearly) vanishing density on alternate
sites and that the other states are (almost) built on either the $n=2$
or $n=3$ Fock states. We, therefore, expect the low energy excitation
to be given by an effective spin-1/2 Heisenberg model, where we map
$|3\rangle$ ($|2\rangle$) to $|\uparrow\rangle$ ($| \downarrow
\rangle$), and remove the state with vanishing density. The new chain
has therefore an effective lattice spacing equal to $2a$.  The
effective interaction arises from the virtual hopping of the bosons to
the empty sites. Taking into account the different intermediate
states, one obtains the following effective spin-half Heisenberg
hamiltonian:
\begin{equation}
  H_{\mathrm{eff}}=\sum_i \frac{J_{\mathrm{eff}}}{2}(S^+_iS^-_{i+1}+S^-_iS^-+_{i+1})
  +\lambda_{\mathrm{eff}}S^z_iS^z_{i+1}-B_{\mathrm{eff}}S^z_i,
\end{equation}
where
\begin{equation}
  \begin{aligned}
    J_{\mathrm{eff}}&=-t^2\left(\frac{3}{4V-2U}\right)\\
    \lambda_{\mathrm{eff}}&=2t^2\left(\frac{3}{4V-2U}+\frac{2}{4V-U}
      -\frac{3}{5V-2U}-\frac{2}{3V-U}\right)\\
    B_{\mathrm{eff}}&=2\delta\mu+2t^2\left(\frac{3}{5V-2U}-\frac{2}{3V-U}\right),
 \end{aligned}
\end{equation}
where $\delta\mu=\mu-2U$. $\lambda_{\mathrm{eff}}$ is positive for a
large range of $(V,U)$ values, so that the preceding Hamiltonian
corresponds essentially the AF spin-half Heisenberg model in a
magnetic field (the negative sign of $J_{\mathrm{eff}}$ can be removed
through the mapping $(S_x, S_y)\rightarrow -(S_x, S_y)$).

Defining the ratio $\rho=2V/U$, one obtains:
\begin{equation}
  \begin{aligned}
    J_{\mathrm{eff}}&=-\frac{t^2}{U}\left(\frac{3}{2\rho-2}\right)\\
    \lambda_{\mathrm{eff}}&=\frac{2t^2}{U}\left(\frac{3}{2\rho-2}+\frac{2}{2\rho-1}
      -\frac{6}{5\rho-4}-\frac{4}{3\rho-2}\right)\\
    B_{\mathrm{eff}}&=2\delta\mu+4t^2\left(\frac{3}{5\rho-4}-\frac{2}{3\rho-2}\right).
 \end{aligned}
\end{equation}
For $B_{\mathrm{eff}}=0$, the system is in the $AF$ ($XY$) phase when
$\lambda_{\mathrm{eff}}<|J_{\mathrm{eff}}|$, ($\lambda_{\mathrm{eff}}
> |J_{\mathrm{eff}}|$)~\cite{1dmagnetism,Giamarchibook}. From the
preceding expressions, the ratio $\Delta = \lambda_{\mathrm{eff}} /
|J_{\mathrm{eff}}|$ starts from the value $\Delta=2$ at $\rho=1$ and
then decreases. The isotropic point $\Delta=1$ is crossed around
$\rho\approx1.15$, such that for $V=0.75U$, i.e. $\rho=1.5$, the
system is the $XY$ phase. The ground state has therefore a vanishing
magnetization, corresponding to an average density $n=5/4$. For $t=0$,
this corresponds to $\delta\mu=0$, i.e.  to the boundary between the
$n=1$ and $n=3/2$ CDW.  Then, for any finite $B_{\mathrm{eff}}$, the
average magnetization is positive (negative) corresponding to an
average density larger (less) than $5/4$.  In addition, the effective
spin correlations exhibit spatial oscillations whose period depends on
the magnetization, which, in turn, leads to gapless excitations at
finite momentum. More precisely, starting from the correlation
functions of the spin-1/2 chain obtained using the bosonisation
approach~\cite{Giamarchibook} and taking into account that
$J_{\mathrm{eff}}$ is negative and that the lattice spacing is $2a$,
one can show that both the in-plane ($\langle S^+S^-\rangle$) and the
out-of-plane ($\langle S^zS^z\rangle$) correlations yield oscillations
corresponding to a wavevector $k_0a=\pi(m+1/2)$, where $m$ is the
magnetization. The latter is related to the average density $\bar{n}$
of the Bose-Hubbard model: $m=2\bar{n}-5/2$, such that
$m+1/2=2(\bar{n}-1)=2\delta n$, where $\delta n=\bar{n}-1$. Therefore,
the gapless excitations correspond to $k_0a=2\pi\delta n$, in perfect
agreement with Figs.~\ref{SS1}, \ref{SS2}, \ref{SS3} and \ref{SS4}.

Finally, one can see in Figs.~\ref{SS1}, \ref{SS2}, \ref{SS3} and
\ref{SS4} that the effective period of the lowest part of the
excitation spectrum is $\pi/2a$, corresponding the doubling of the
lattice spacing. This is not true for the higher excitations which are
gapped and most likely involving the empty sites.

\subsection{SS-SF transition}

We have also studied the evolution of the structure factor
$S(k,\omega)$ across the supersolid-superfluid transition, at fixed
density and interaction strengths, increasing the hopping amplitude
from $t/U=0.24$ (SS) to $t/U=0.3$ (SF).

\subsubsection{TEBD results}

\begin{figure}[H]
  \centerline{\includegraphics[width=6cm]{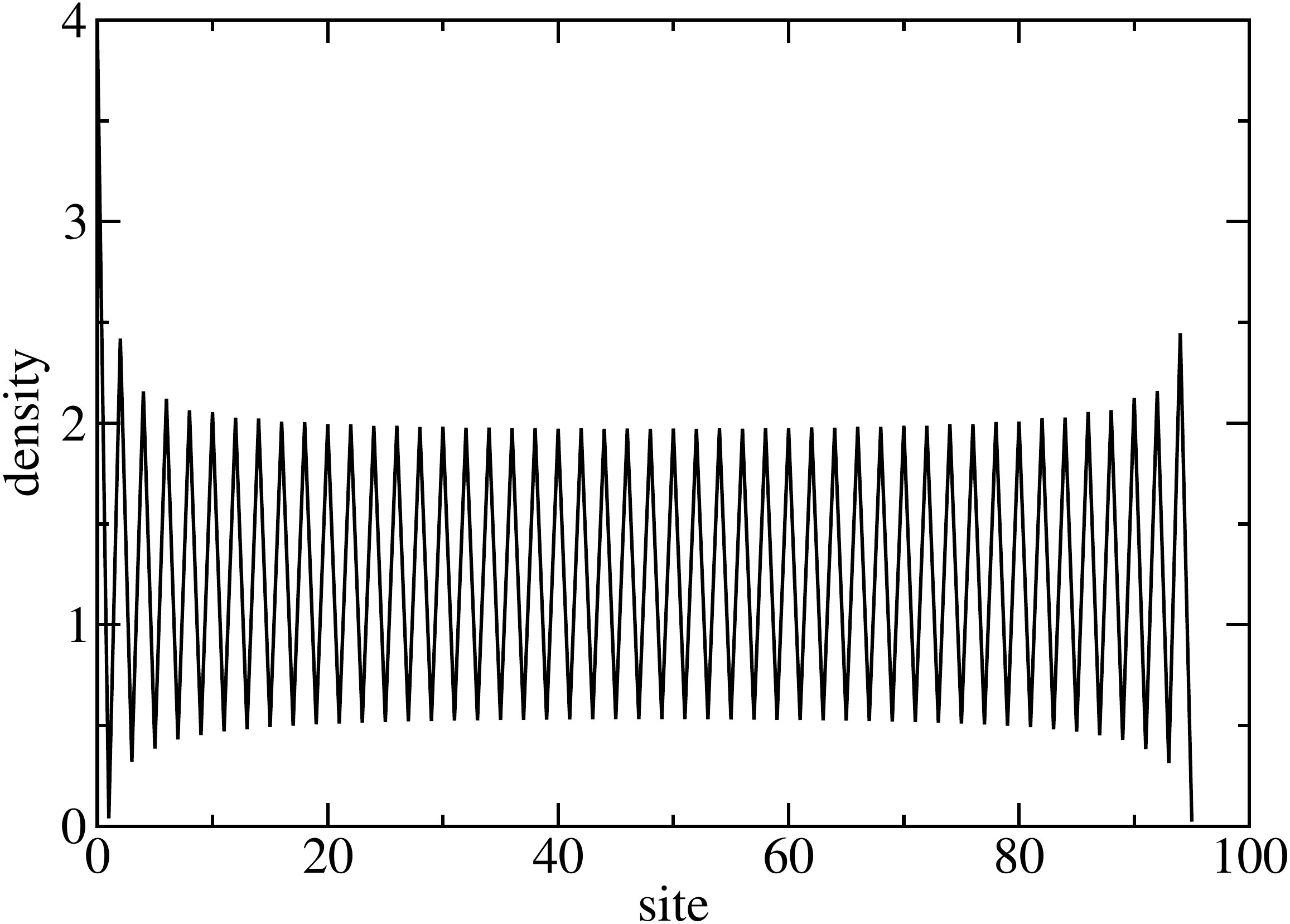}}
  \centerline{\includegraphics[width=6cm]{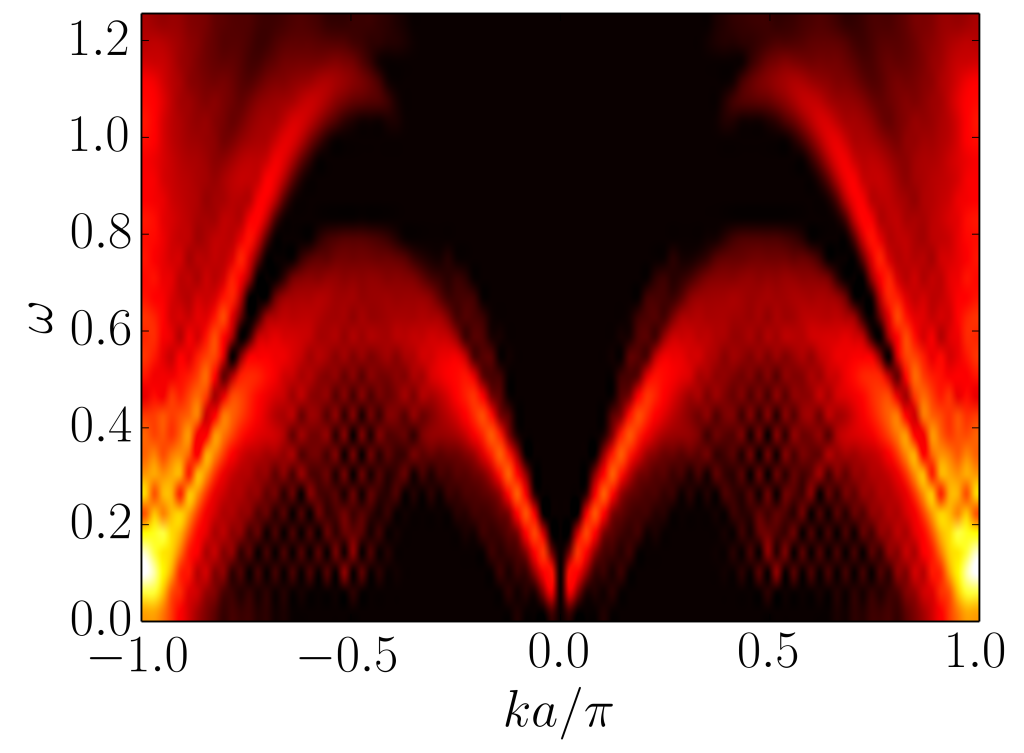}}
  \caption{\label{SS_t_0.24_profile}The SS phase at $U=1$, $V=0.75$
    and $t=0.24$, $n=1.25$. Top: Density profile: The CDW order is
    still almost perfect. Bottom: The gapless modes at $ka=0.5\pi$ are
    still visible, but their contributions to $S(k,\omega)$ have a
    smaller weight when compared with Fig.~\ref{SS1}.}
\end{figure}

\begin{figure}[H]
  \centerline{\includegraphics[width=6cm]{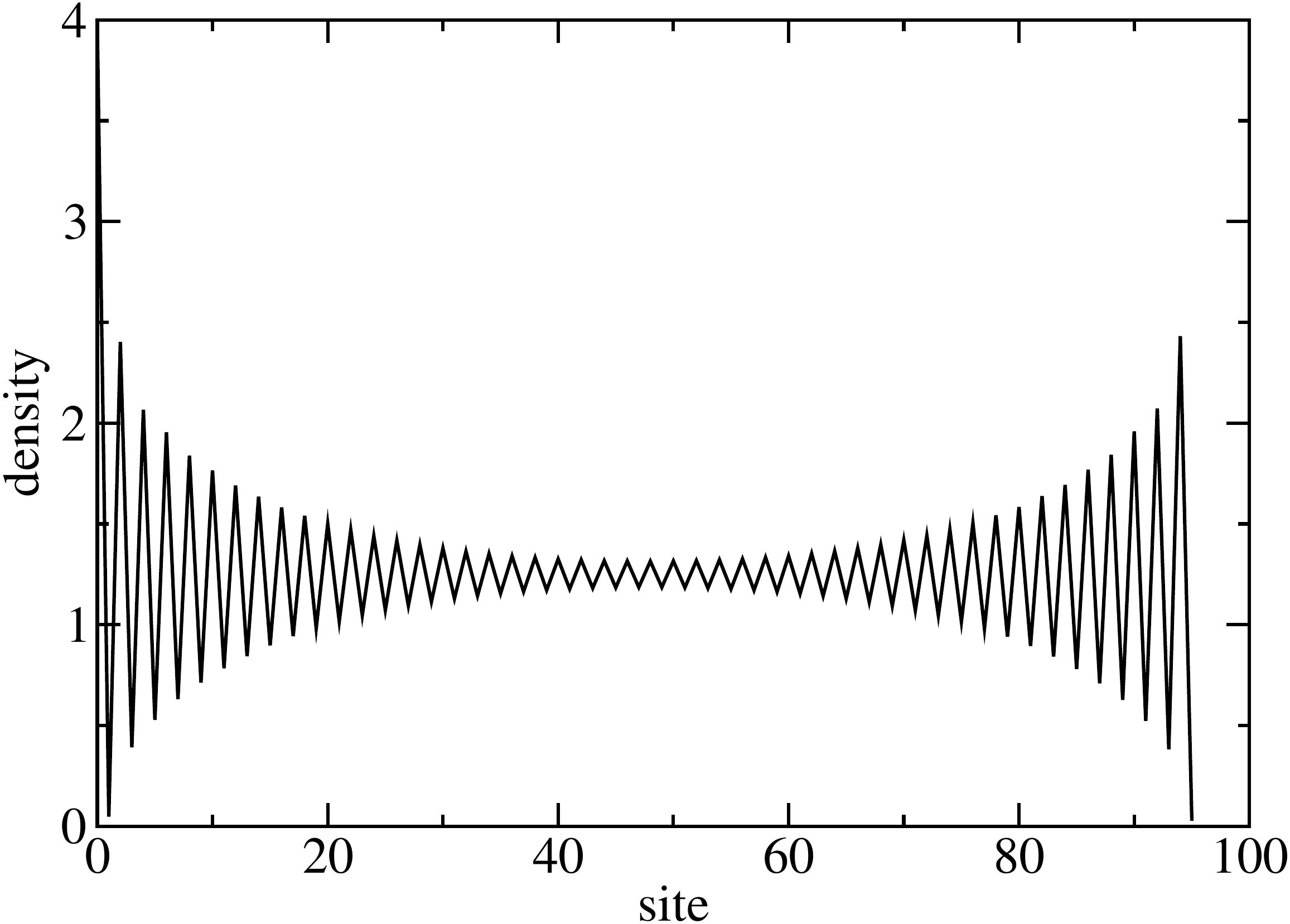}}
  \centerline{\includegraphics[width=6cm]{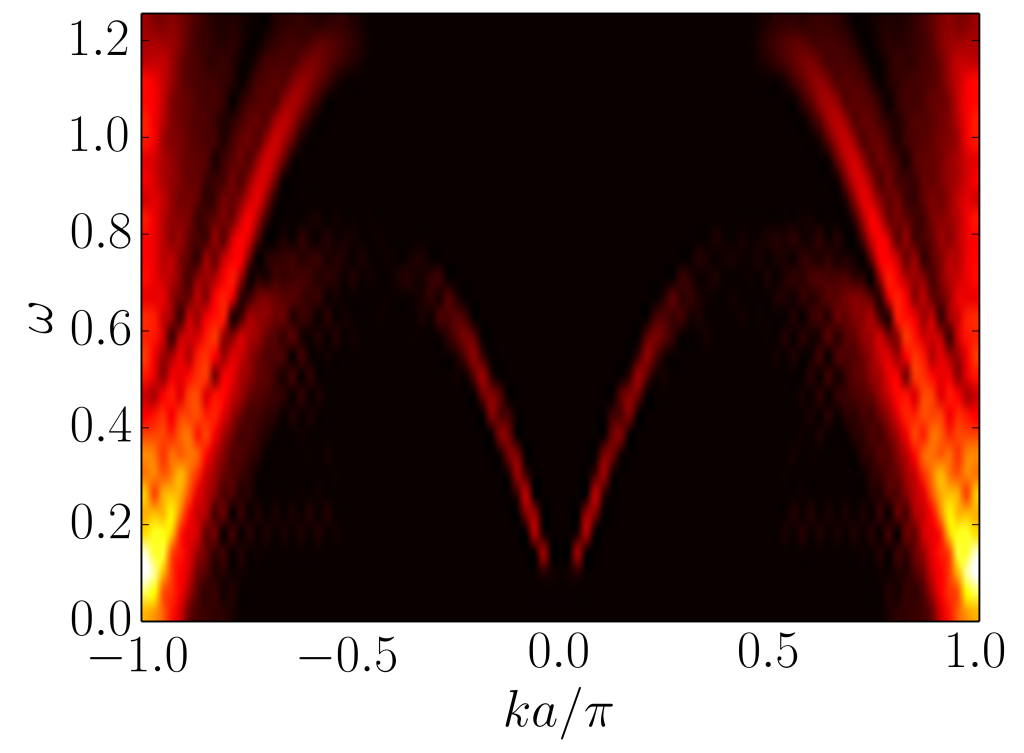}}
  \caption{\label{SS_t_0.26_profile}The supersolid phase at $U=1$,
    $V=0.75$ and $t=0.26$, $n=1.25$.  Top: The density pattern does
    not show a well defined CDW. Bottom: The gapless mode at
    $ka=0.5\pi$ has almost disappeared, but the system is still
    gapless at $ka=\pi$}
\end{figure}

\begin{figure}[H]
  \centerline{\includegraphics[width=6cm]{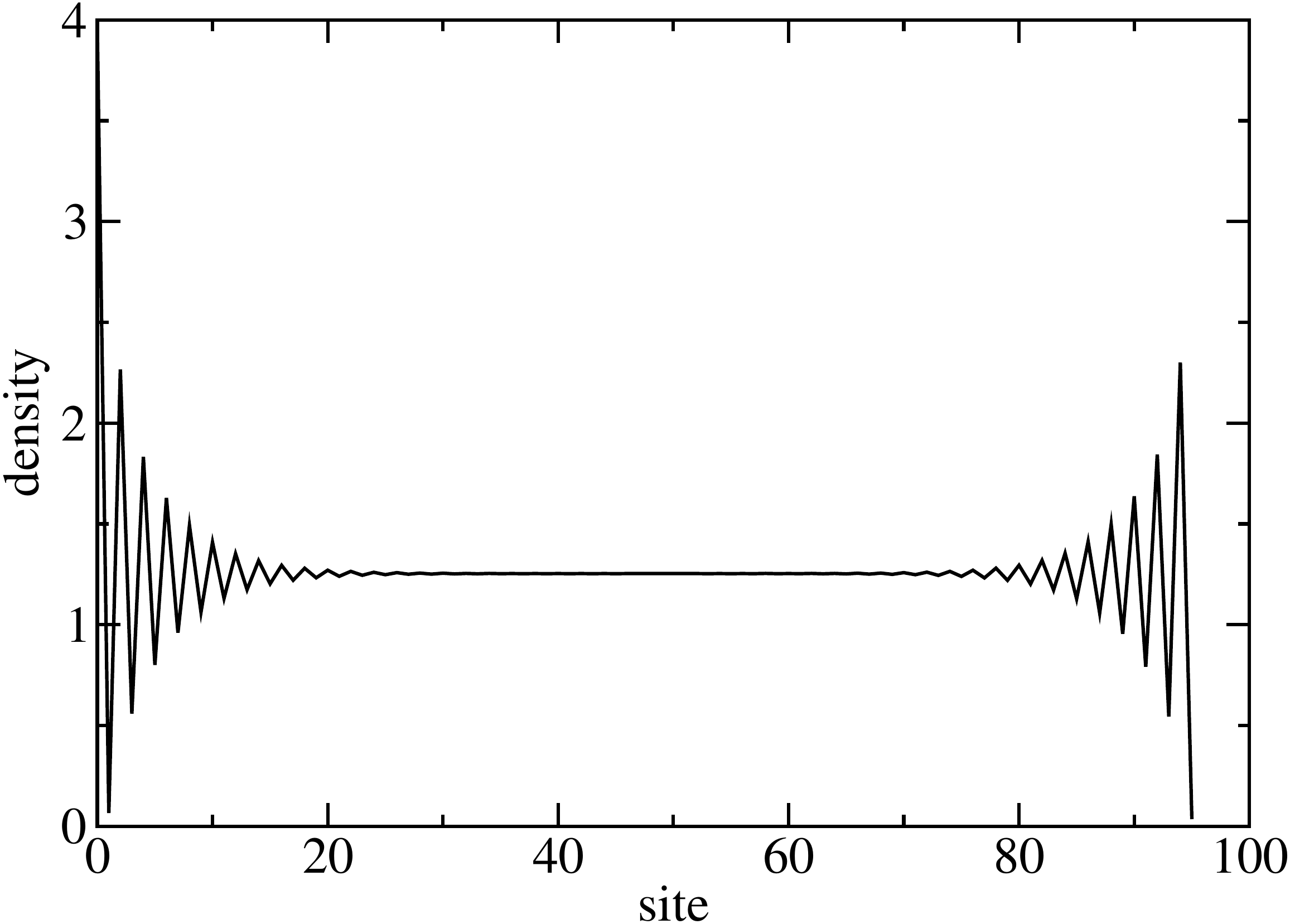}}
  \centerline{\includegraphics[width=6cm]{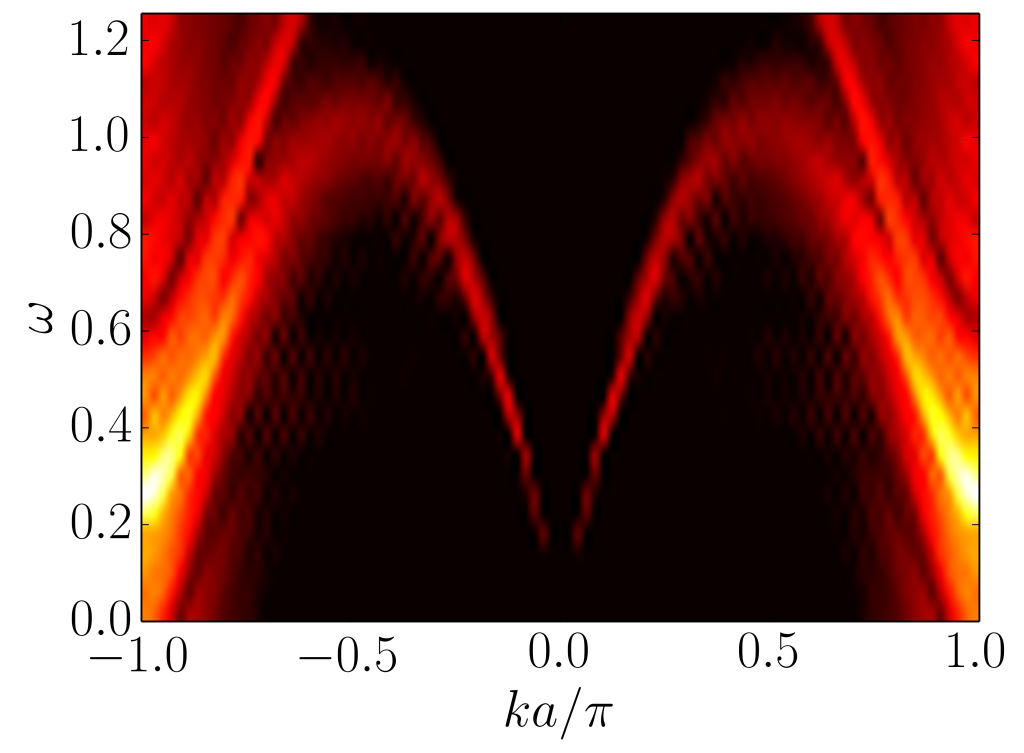}}
  \caption{\label{SS_t_0.3_profile} The superfluid phase at $U=1$,
    $V=0.75$ and $t=0.3$, $n=1.25$. Top: The density profile no longer
    shows CDW pattern.  Bottom: The only gapless mode is at $ka=0$, as
    expected in the SF phase; at $ka=\pi$, the system is now gapped, }
\end{figure}

From the density plots, see Fig.~\ref{SS_t_0.24_profile} (top),
Fig.~\ref{SS_t_0.26_profile} (top) and Fig.~\ref{SS_t_0.3_profile}
(top), we see that the SS-SF transition is driven by the disappearance
of the spatial modulation. This behavior is also predicted by the
standard mean-field theory where the ground state of the system is
assumed to be a tensor product of onsite wavefunctions (see below).

The disappearance of the spatial modulation results in an opening of
the gap at $ka=\pi$, see Fig.~\ref{SS_t_0.3_profile} (bottom), which
is well described by the mean-field theory, since it only amounts to a
change of the spatial periodicity, i.e. from $2a$ to $a$, of the
effective Hamiltonian~\cite{Iskin}.

On the other hand this simple mean-field cannot capture the long range
quantum correlations that lead to the gapless modes at
$ka=k_{SS}a=2\pi\delta n$ and the mapping to the spin-1/2 is no longer
valid close to the transition since one cannot neglect previously
empty sites.  From that point of view, the exact fate of these gapless
mode is still lacking a physical explanation.

\subsubsection{Mean-field results}

As explained above, a well known mean-field method to solve the
Bose-Hubbard model is the Gutzwiller ansatz, where the ground state
wavefunction is assumed to be a tensor product of onsite
wavefunctions:
\begin{equation}
  |\Psi \rangle = \bigotimes_i |\psi_i\rangle\text{ where }
  |\psi_i\rangle=\sum_{n=0}^{N_{\max}} f_{n,i} |n,i\rangle.
\label{eq:Gutzwiller}
\end{equation}
$|n,i\rangle$ represents the Fock state of $n$ atoms occupying the
site $i$, $n_{\max}$ is a cut off in the maximum number of atoms per
site, and $f_{n,i}$ is the probability amplitude of having the site
$i$ occupied by $n$ atoms.

Minimizing the mean-field energy $\langle \Psi| H |\Psi\rangle$ over
the $f_{n,i}$ allows us to determine the mean-field ground state
properties as functions of the different parameters $(U,t,V,\mu)$. For
instance, the superfluid phase corresponds to a non vanishing value of
the order parameter $\langle\Psi|b|\Psi\rangle$, whereas the Mott
phase corresponds to a vanishing order parameter and the $\psi_i$ are
pure Fock states.  In the CDW phase, the order parameter $\langle
b\rangle$ vanishes; the density, $\langle n(ka=0)\rangle$ and the
staggered density, $\langle n(ka=\pi)\rangle$, have the same value.
The supersolid phase corresponds to non-vanishing values for both
$\langle b(ka=0)\rangle$ and $\langle b(ka=\pi)\rangle$; the density
still exhibits oscillations at $ka=\pi$. The superfluid phase
corresponds to a homogeneous density and only the $ka=0$ order
parameter $\langle b(ka=0)\rangle$ has a non-vanishing value.

\begin{figure}[H]
  \centerline{\includegraphics[width=8cm]{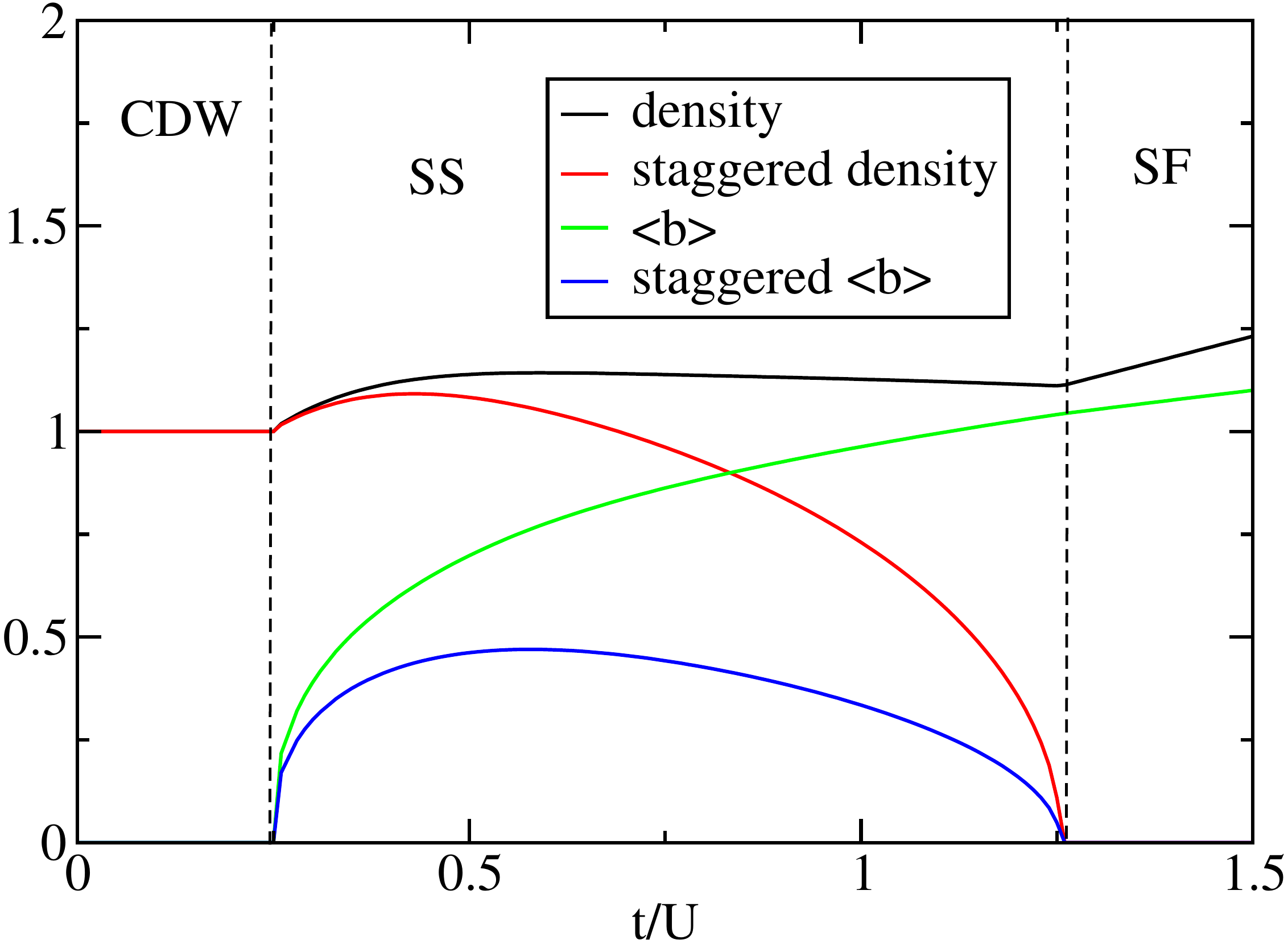}}
  \caption{\label{mfphase}(Color online) Mean field phase diagram for
    $U=1$, $V=1.5$ and $\mu=1.8$ as a function of $t$.  For $0\leq
    t\leq 0.25$, the system is in a CDW insulating phase: the order
    parameter $\langle b\rangle$ vanishes and the density and
    staggered density have the same value.  The CDW corresponds to a
    density pattern $\cdots2020202020\cdots$.  For $0.25\leq t\leq
    1.26$ the system is in the supersolid phase: both $\langle
    b(k=0)\rangle$ and $\langle b(ka=\pi)\rangle$ are
    non-vanishing. The density still presents oscillations at
    $ka=\pi$.  For $t\geq1.25$, the system is superfluid. }
\end{figure}

We present mean-field results for $U=1$, $V=1.5$ and $\mu=1.8$. Note
that since the chemical potential is fixed, the density changes as
$t/U$ is changed. Figure~\ref{mfphase} shows the different quantities
as functions of $t/U$. For $0\leq t\leq 0.25$, the system is in the
CDW insulating phase: The order parameter $\langle b\rangle$ vanishes
and the density and staggered density have the same value; the CDW
corresponds to a density pattern $\cdots2020202020\cdots$.  For
$0.25\leq t\leq 1.26$ the system is in the supersolid phase: Both
$\langle b(k=0)\rangle$ and $\langle b(ka=\pi)\rangle$ are
non-vanishing. The density still presents oscillations at $ka=\pi$.
For $t\geq1.25$, the system is in the usual superfluid phase.
 
The mean field approach also allows us to compute the excitation
spectrum. Since in both the CDW and SS phases the periodicity of the
ground state is $2a$, the spectrum is defined in the reduced Brillouin
zone $[-\pi/2a,\pi/2a]$ and has two branches. In the CDW phase, the
elementary excitations are gapped, as expected. In the SS phase, the
lower branch becomes gapless with a linear behavior around $k=0$. At
the SS-SF transition, the periodicity of the ground state goes back to
$a$, so that the two elementary excitation branches merge at
$ka=\pi/2$. 

\section{underdoped half-filling CDW}
\label{sec:undercdw}

\begin{figure}[H]
  \centerline{\includegraphics[width=8cm]{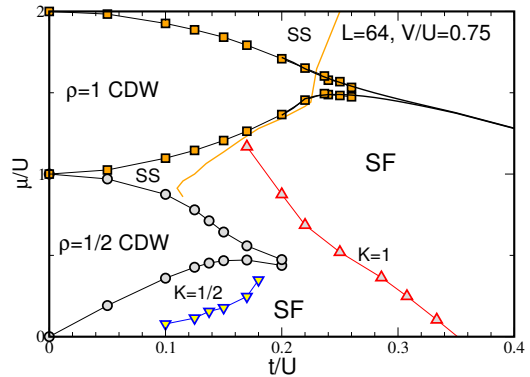}}
  \caption{\label{Phase_diag_details}(color online) Detail of the
    $n=1/2$ lobe (from QMC) where we also determined the constant $K$
    lines for $K=1$, $1/2$.}
\end{figure}

\begin{figure}[H]
  \centerline{\includegraphics[width=8cm]{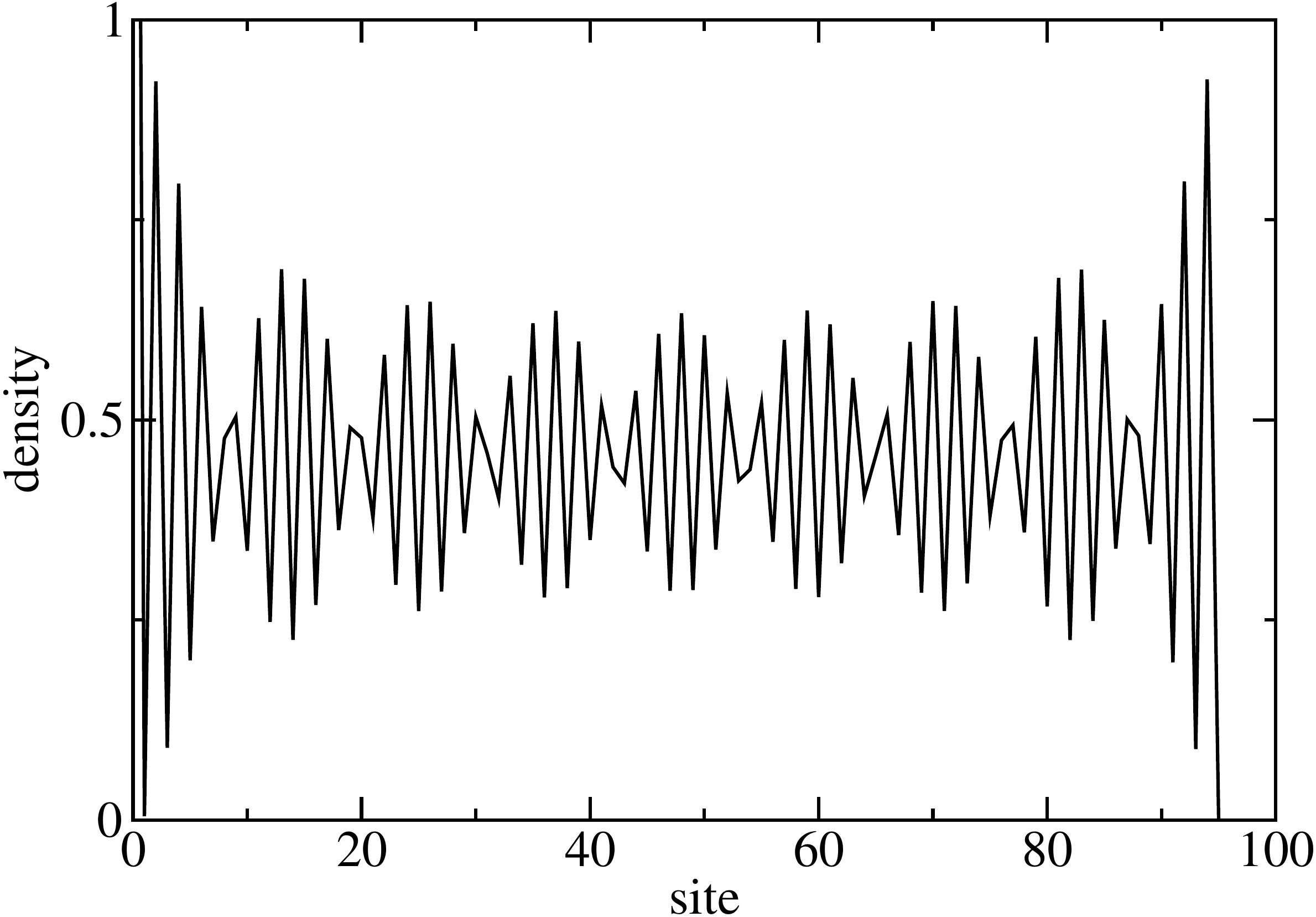}}
  \caption{\label{UCDW}Density profile in the underdoped 1/2 CDW phase
    $U=1$, $V=0.75$ and $t=0.1$, $n=0.4375$. The overall modulation of
    the density emphasizes the difference with the SS phase.}
\end{figure}

\begin{figure}[H]
  \centerline{\includegraphics[width=4cm]{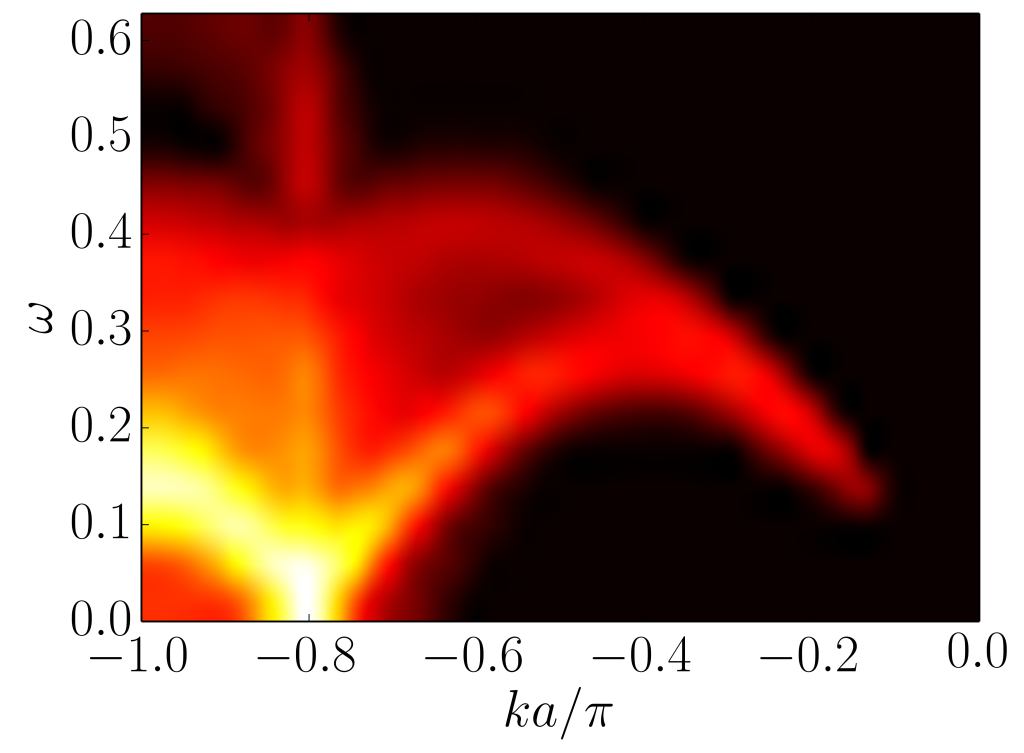}
    \includegraphics[width=4cm]{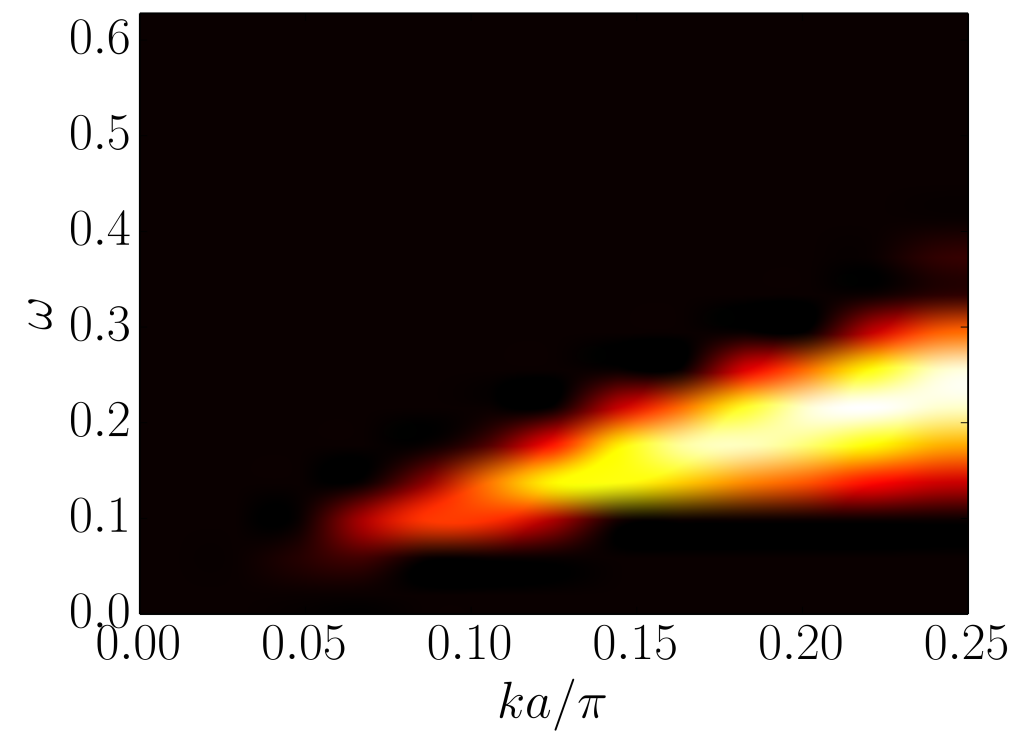}}
  \caption{\label{UCDW_SF}(color online) Structure factor in the
    underdoped CDW: $U=1$, $V=0.75$ $t=0.2$, the density is
    $n=0.40625$. The gapless mode at $ka=0$ indicates the ODLRO. The
    gapless mode at $ka\approx0.8\pi$ is compatible with the
    bosonisation prediction $2\pi n$ for an AF spin-1/2 chain in a
    finite magnetic field.}
\end{figure}

In this section, we compare the structure factor obtained in the SS
phase with the one for the phase between the half-filled CDW and the
superfluid phase, see Fig.~\ref{Phase_diag_details}. A typical density
profile is shown in Fig.~\ref{UCDW}; the parameters are $U=1$,
$V=0.75$ and $t=0.1$, corresponding to an average density
$n=0.4375$. One clearly sees that the density pattern is different
from the one in the supersolid phase: the long wavelength modulation
of the CDW is a signature of a vanishing DLRO, in contrast to the SS
phase. On the other hand, one has an overall power law decay of the
ODLRO $g(x)\propto x^{-1/2K}$, but with a coefficient $K<1/2$
emphasizing that the SF can be localized with a single
impurity~\cite{kuhner00}.  The difference with the SS phase also
appears in the structure factor, Fig.~\ref{UCDW_SF}: One has only two
gapless modes, one at $ka=0$ and one at $ka\approx0.8\pi$, but the
excitations at $k=\pi$ are gapped. Therefore, the periodicity of the
lower part of the spectrum is just $2\pi/a$ and not $\pi/a$ as in the
SS phase.

Since, at very low values of $t$, the density pattern for the
half-filled CDW phase is $\cdots01010101\cdots$, there is a natural
mapping onto a spin-half AF Heisenberg model with a vanishing total
magnetization: $|0\rangle\rightarrow |\downarrow\rangle$ and
$|1\rangle\rightarrow |\uparrow\rangle$.  The underdoped CDW phase
corresponds then to a non vanishing total magnetization $S_z\approx
n-1/2$. However, contrary to the SS phase, there is no simple way to
get the effective $J$ and $\lambda$ coefficients: the initial state
$\cdots01000101\cdots$ and a state after one hopping
$\cdots01001001\cdots$ are actually degenerate in the limit
$t\rightarrow0$, thereby preventing a proper expansion of the
Bose-Hubbard Hamiltonian. Nevertheless, one can still argue that the
gapless mode at $ka\approx0.8\pi$ corresponds to the incommensurate
gapless mode appearing in the spin-spin correlation function for the
spin-half AF Heisenberg model in a finite magnetic field. In
particular the value $k_0a=0.8\pi$ is compatible with the bosonisation
prediction $2\pi n$~\cite{Giamarchibook}.

\section{Conclusions}
\label{sec:conclusion}

In summary, we have studied the excitation spectra of the extended
Bose-Hubbard model. Along, the MI-HI-CDW transition, the dynamical
structure factor exhibits behavior similar to the spin-spin
correlation for the $S=1$ Heisenberg model. For instance, it shows a
difference between the neutral and charge gaps in the HI phase.

In the SS phase, we have shown that the system has additional gapless
modes at a $k$ value that depends on the average density of the
system. They can be mapped to the incommensurate gapless modes of an
AF spin-1/2 chain at finite magnetization.  They are a signature of
the DLRO present in the SS phase. These modes fade away when moving
towards the SF phase, and, in addition, a gapped mode appears at
$ka=\pi$, marking the disappearance of the DLRO.

Finally, we have shown that underdoping the CDW at half-filling, the
excitation spectrum differs from the one in the SS phase, emphasizing
that even though the system exhibits superfluidity and oscillations in
the density, there is no DLRO.

\bigskip

\begin{acknowledgments}

The Centre for Quantum Technologies is a Research Centre of Excellence funded by
the Ministry of Education and National Research Foundation of Singapore.

\end{acknowledgments}

\end{document}